\documentclass[eqsecnum,aps,showpacs,amsmath,nofootinbib,superscriptaddress,preprintnumbers]{revtex4}

\usepackage{epsfig, color,graphicx}
\usepackage{amsmath}
\usepackage{amssymb}
\newcommand{\nn}{\nonumber \\}
\newcommand{\bea}{\begin{eqnarray}}
\newcommand{\ena}{\end{eqnarray}}
\newcommand{\vs}[1]{\vspace{#1 mm}}
\newcommand{\hs}[1]{\hspace{#1 mm}}
\renewcommand{\a}{\alpha}

\renewcommand{\c}{\gamma}
\renewcommand{\d}{\delta}
\newcommand{\e}{\epsilon}
\newcommand{\s}{\sigma}

\newcommand{\la}{\lambda}
\newcommand{\pa}{\partial}
\newcommand{\p}[1]{(\ref{#1})}

\newcommand{\lsim}{\, \mbox{\raisebox{-1.ex}
{$\stackrel{\textstyle<}{\textstyle\sim}$}}\,}

\newcommand{\tr}{\tilde{r}}

\newcommand{\squaret}{\kern1pt\vbox{\hrule height 0.9pt\hbox{\vrule width
0.9pt\hskip 2pt\vbox{\vskip 5.5pt}\hskip 3pt\vrule width 0.3pt}\hrule height
0.3pt}\kern1pt}
\begin{document}

\preprint{KU-TP 057}
%%
%<<<<<<<<<<<<< TITLE >>>>>>>>>>>>>>>%
%%
\title{Charged Black Holes in String Theory with Gauss-Bonnet Correction
in Various Dimensions
}
%%
%<<<<<<<<<<<<< AUTHOR >>>>>>>>>>>>>>>%
%%
\author{Nobuyoshi Ohta}\email{ohtan@phys.kindai.ac.jp}
\affiliation{Department of Physics, Kinki University, Higashi-Osaka,
Osaka 577-8502, Japan}

\author{Takashi Torii}\email{torii@ge.oit.ac.jp}
\affiliation{Department of General Education, Osaka Institute of Technology,
Asahi-ku, Osaka 535-8585, Japan}

%<<<<<<<<<<<<< DATE >>>>>>>>>>>>>>>%
\date{\today}

%======================================%
%<<<<<<<<<<<<< ABSTRACT >>>>>>>>>>>>>>>%
%======================================%
\begin{abstract}
We study charged black hole solutions in Einstein-Gauss-Bonnet theory with the dilaton
field which is the low-energy effective theory of the heterotic string.
The spacetime is $D$-dimensional and assumed to be static and spherically symmetric with
the $(D-2)$-dimensional constant curvature space and asymptotically flat.
The system of the basic equations is complex and the solutions are obtained numerically.
We identify the allowed parameter region where the black hole solutions exist, and show
configurations of the field functions in  $D=4$ -- 6 and 10. We also show the relations
of the physical quantities of the black holes such as the horizon radius, the mass,
the temperature, and so on, and find several results.
The forms of the allowed parameter regions are different depending on the dimension.
There is no extreme black hole solution with $T=0$ that can be obtained by taking the
limit of the non-extreme solutions within the parameter range we chose.
Entropy of the black holes in the dilatonic theory is always larger than that in
the non-dilatonic theory.
Our analysis includes the higher order term of the dilaton field which is not in our
previous works.
Its effect remarkably appears in five dimensions and is given in the appendix.

By our analysis it is found that
the properties of the black hole solutions strongly depend on the dimension,
charge, existence of the dilaton field. Hence both the detailed analyses of the
individual systems and the investigations from the systematic point of view
are important.
\end{abstract}

%<<<<<<<<<<<<< PACS NUMBER >>>>>>>>>>>>>>>%
\pacs{04.60.Cf, 04.50.Gh, 04.50.-h, 11.25.-w }

% 04.50.Gh : higher-dimensional Black holes
% 04.50.-h : Higher-dimensional gravity and other theories of gravity
% 04.60.Cf : gravitational aspects of String theory
% 11.25.-w : Strings and branes

\maketitle

%======================================%
%======================================%
%<<<<<<<<<<<< SECTION I  >>>>>>>>>>>>>>%
%======================================%
%======================================%
\section{Introduction}
\label{intro}

One of the most important problems in theoretical physics is to find
the quantum theory of gravity and to apply it to physical systems to
understand physics at strong gravity such as a black hole singularity.
The leading candidates for that including
all the fundamental forces of elementary particles are the ten-dimensional
superstring theories or eleven-dimensional M-theory. The area where such quantum gravity
plays the significant role includes the cosmology and black hole physics.
There has been interest in applications of string/M theories to these subjects.
Since it is still difficult to study geometrical settings in string/M theories directly,
most analyses have been made by using low-energy effective theories
inspired by string/M theories. The effective theories are the supergravities
which typically involve not only the metric but also the dilaton field
(as well as several gauge fields).

The first attempt at understanding black holes in the Einstein-Maxwell-dilaton
system was made in Refs.~\cite{GM,GHS}, in which a static spherically symmetric
black hole solution with a dilaton hair was found. After that, many solutions
were discussed in various models. On the other hand,
it is known that there are correction terms of higher orders in the curvature to the
lowest effective supergravity action coming from the superstring theories.
The simplest correction is the Gauss-Bonnet (GB) term coupled to the dilaton field
in the low-energy effective heterotic string~\cite{MT}.
It is then natural to ask how the black hole solutions are affected by
the higher order terms in the effective theories.

When the dilaton field is dropped or is set to a constant by hand, the total action consists of
the cosmological constant, Einstein-Hilbert and GB term, which are
the first three terms in the Lovelock theory~\cite{Lovelock,Zumino}.
Motivated by this observation, there have been many works on the black hole
solutions in the Lovelock theory~\cite{GG,CNO,KMR,TYM}.
In the four-dimensional spacetime,
the GB term does not give any contribution because it becomes a surface term
and gives a topological invariant. Boulware and Deser (BD)~\cite{BD}
discovered a static, spherically symmetric black hole solutions of such
models in more than four dimensions. In the system with a negative cosmological
constant, black holes can have horizons with nonspherical topology such as
torus, hyperboloid, and other compactified submanifolds. These
solutions were originally found in general relativity and are
called topological black holes~\cite{bro84}. Topological black
hole solutions were studied also in the Einstein-GB (EGB) theory~\cite{Cai}.
It is very interesting to see how these are modified by the presence of a dilaton field.
One of the purposes of this paper is to study charged black hole
solutions with higher order corrections as well as the dilaton field.

%Another motivation for this work is the following.
There has been recently a renewed interest in these solutions as an application to the
calculation of shear viscosity in the strongly coupled gauge theories using the black hole
solutions in the five-dimensional EGB theory via AdS/CFT
correspondence~\cite{vis,CNOS}.
Almost all these studies considered the pure GB term without the dilaton field,
or assumed a constant dilaton, which is not a solution of the heterotic string theory.
It is, however, expected that AdS/CFT correspondence is valid within
the effective theories of superstring which necessarily involve the dilaton field.
The inclusion of the dilaton field was also considered by Boulware and Deser~\cite{BD},
but exact black hole solutions and their thermodynamic properties
were not discussed. Callan {\it et al.}~\cite{cal88} considered
black hole solutions in the theory with a higher-curvature term
$R_{\mu\nu\rho\sigma}R^{\mu\nu\rho\sigma}$ and the dilaton field,
and Refs.~\cite{KMR,TYM} took both the GB term and the dilaton
field into account in four-dimensional spacetime.

Hence it is important to study black hole solutions and their properties
in the theory with both the higher order corrections and the dilaton field in general dimensions.
In our earlier paper~\cite{GOT1}, we focused on asymptotically
flat solutions and studied a system with the GB correction term and the dilaton field without
the cosmological constant in various dimensions from 4 to 10.
They are spherically symmetric with the $(D-2)$-dimensional
hypersurface of curvature signature $k=1$.
We have then turned to planer symmetric ($k=0$) topological black holes,
but have found that no solution exists without the cosmological constant.
In the string perspective, it is more interesting to examine asymptotically
anti-de Sitter (AdS) black hole solutions with possible application to the AdS/CFT
correspondence in mind.
So in the sequel paper~\cite{GOT2}, we have presented the results
on black hole solutions with a negative cosmological constant with $k=0$.
In fact, shear viscosity has been computed using the result in that paper~\cite{CNOS}.
Other cases of solutions with $k=\pm 1$ and a negative cosmological
term as well as topological black holes with and without the cosmological constant were
studied in \cite{OT3,OT4}, and their global structures were studied in Ref.~\cite{OT5}.
Cosmological solutions were also considered in Ref.~\cite{BGO}.
Other solutions are discussed in Refs.~\cite{TM,CGO1,CGO2,CCMO,MOS1,CGOO,MOS2,CGK}.

In the study of superconductors and superfluidity using the AdS/CFT correspondence,
charged black hole solutions play important roles~\cite{Gubser}.
Hence it is also significant to extend our above studies of the neutral black hole solutions
to charged ones. In particular, we should consider the inclusion of the dilaton field again.

There is another reason for our study of the system.
In our above study of the neutral black holes~\cite{GOT1,GOT2,OT3,OT4,OT5},
we did not consider the higher derivative term of the dilaton field, which also appears
naturally as we will see below~\cite{MT}.
We intend to incorporate this term also and study how this modifies the solutions.
As a result of our analysis, however, there is not much qualitative
difference. Thus the results in our earlier papers should be useful.

This paper is organized as follows. In Sec.~\ref{sec2}, we first present the action
and give basic equations to solve for the system of Einstein-Maxwell-Gauss-Bonnet term
coupled to the dilaton field with a cosmological constant, although we will focus on the
asymptotically flat solutions without cosmological constant later on.
The case with the cosmological constant is left for future study.
Boundary conditions and symmetry properties of the theory are also discussed in order
to apply them in our following analysis.
In Sec.~\ref{sec3}, we briefly summarize the thermodynamic properties of the solutions
in the theory.
In Sec.~\ref{sec4}, we also briefly review the results for non-dilatonic solutions
to see the difference from the dilatonic ones.
We then proceed to the study of the charged black holes for $D=4,5,6$ and 10
in Secs.~\ref{secd=4}, \ref{secd=5}, \ref{secd=6} and \ref{secd=10}, respectively.
The study of the effects of the higher derivative of the dilaton field for the neutral
black holes is delegated to the appendix, because we find that there is not much
difference in the properties of the black hole solutions except for five dimensions.
Some detailed discussions are given there.
Section~\ref{CD} is devoted to conclusions and discussions.

%%%%%%%%%%%%%%%%%%%%%%%%%%%%%%%%%%%%
%%%%%%%%%%%%%%%%%%%%%%%%%%%%%%%%%%%%
\section{Dilatonic Einstein-Maxwell-Gauss-Bonnet theory}
\label{sec2}
%%%%%%%%%%%%%%%%%%%%%%%%%%%%%%%%%%%%
%%%%%%%%%%%%%%%%%%%%%%%%%%%%%%%%%%%%

%%%%%%%%%%%%%%%%%%%%%%%%%%%%%%%%%%%%
\subsection{Action and basic equations}
%%%%%%%%%%%%%%%%%%%%%%%%%%%%%%%%%%%%

We consider the following low-energy effective action for the
heterotic string theory in one scheme~\cite{MT}:
\bea
S &=& \frac{1}{2\kappa_D^2}\int d^{D}x \sqrt{-g} e^{-2\phi} \Bigg[R +4 (\pa \phi)^2
-\frac{1}{4}F^2 -\frac{1}{12} H^2 + \a_2 \Big\{ R^2_{\rm GB} -16 \Big(R^{\mu\nu}
-\frac{1}{2} R g^{\mu\nu} \Big)
\pa_\mu \phi \pa_\nu \phi \nn
&& + 16 \squaret \phi(\pa\phi)^2 - 16(\pa\phi)^4
-\frac12 \Big( R^{\mu\nu\rho\s}H_{\mu\nu\a}H_{\rho\s}{}^\a-2R^{\mu\nu}H_{\mu\nu}^2
+\frac13 RH^2 \Big) + 2\Big(D^\mu\pa^\nu \phi H_{\mu\nu}^2-\frac13 \squaret\phi H^2\Big) \nn
&& +\frac23 H^2(\pa\phi)^2 + \frac{1}{24}H_{\mu\nu\la}H^\nu{}_{\rho\a}H^{\rho\s\la}
H_\s{}^{\mu\a}
- \frac18 H_{\mu\nu}^2 H^{2\mu\nu} + \frac{1}{144}(H^2)^2
\Big\}\Bigg],
\label{action0}
\ena
where $\kappa_{10}^2$ is a $D$-dimensional gravitational constant,
$\phi$ a dilaton field, $F$ a gauge field strength, $H$ a 3-form, $\alpha_2=\a'/8$ is a numerical
coefficient given in terms of the Regge slope parameter $\a'$, and
$R^2_{\rm GB} = R_{\mu\nu\rho\sigma} R^{\mu\nu\rho\sigma}
- 4 R_{\mu\nu} R^{\mu\nu} + R^2$
is the GB term. $H^2=H^{\mu\nu\rho}H_{\mu\nu\rho}$ and
$H_{\mu\nu}^2=H_\mu{}^{\rho\s}H_{\nu\rho\s}$.
In the original derivation of the effective action, it was first derived in the
Einstein frame from the S-matrix calculation in the string theory, and then transformed
into the string frame~\cite{MT}. It is common and convenient to interpret results
in the Einstein frame. Hence we transform Eq.~(\ref{action0}) into the Einstein frame, reduce to $D$
dimensions, and use the field redefinition ambiguity~\cite{MT,MOW}
\bea
\d g_{\mu\nu}&=&\a'[a_1 R_{\mu\nu}+a_2 \nabla_\mu\phi\nabla_\nu \phi+g_{\mu\nu}\{a_3 R
+a_4 (\nabla\phi)^2 +a_5 \nabla^2\phi\}], \\
\d\phi&=&\a'[b_1 R+ b_2(\pa\phi)^2+b_3 \nabla^2\phi],
\ena
to obtain, up to higher order terms,
\bea
S &=& \frac{1}{2\kappa_D^2}\int d^{D}x \sqrt{-g} \Bigg[R -\frac12 (\pa \phi)^2
-\frac{1}{4}e^{-\c\phi} F^2 + \a_2 e^{-\c\phi} \Big\{ R^2_{\rm GB}
+ \frac{3}{16} a (\pa\phi)^4 \Big\} -\Lambda e^{\la\phi} \Bigg],
\label{action1}
\ena
where $\c=1/2$. We have set $H=0$ because we focus on the effects of the
dilaton field and the gauge field.
Note that $H=0$ is a solution of the field equations.
%so this procedure is legitimate.

When the action~\p{action1} is derived from \p{action0}, a numerical constant $a$
in the action~\p{action1}  should be $a=1$.
In our previous papers~\cite{GOT1,GOT2,OT3,OT4,OT5}, however,
we have examined the system with $a=0$, which corresponds to the case without
the higher order term of the dilaton field.
To see how the solutions are different with and without the higher order term of
the dilaton field, we have included the constant $a$, and here we study mainly $a=1$ case.
The difference between $a=0$ and $a=1$ is discussed in the Appendix.
We have also included a cosmological constant with the dilaton coupling $\la$
although we study asymptotically flat solutions in this paper.

In the above process, higher order terms ($\geq O(\alpha'^{2})$) are dropped.
This is allowed because the effective low-energy action can be determined up to the
field redefinition when it is read off from the scattering amplitudes computed
in the string theories.
Also note that the effective action was originally computed in the Einstein frame.
All this means that there is no absolutely preferred form of the action if they
are related up to terms of order $\alpha'^2$.
There might be some quantitative differences if we adopt theory of different choice
of the coefficients, but we expect that the qualitative properties will not
change significantly. It is the system~\p{action1} with $a=1$ that we study in this paper.

Let us consider the metric and field strength
\bea
ds_D^2 = - B(r) e^{-2\d(r)} dt^2 + B(r)^{-1} dr^2 + r^2 h_{ij}dx^i dx^j,~~
F_{0r}=\frac{df(r)}{dr},
\ena
where $h_{ij}dx^i dx^j$ represents the line element of a
$(D-2)$-dimensional constant curvature space with curvature
$(D-2)(D-3)k$ and volume $\Sigma_k$ for $k=\pm 1,0$.
%$\Sigma_1= \frac{2 \pi^{(D-1)/2}}{\Gamma((D-1)/2)}$

The field equations following from Eq.~\p{action1} are
\bea
&& \bigl[(k-B)\tr^{D-3}\bigr]' \frac{D-2}{\tr^{D-4}}h -\frac12 B \tr^2 {\phi'}^2
 - (D-1)_4\,e^{-\c\phi}\frac{(k-B)^2}{\tr^2}
 + 4(D-2)_3\, \c e^{-\c\phi}B(k-B)(\phi''-\c {\phi'}^2) \nn
&& \hs{5} + 2(D-2)_3\,\c e^{-\c\phi}\phi'\frac{(k-B)[(D-3)k-(D-1)B]}{\tr}
-\frac{\tr^2}{2} e^{2\d-\c\phi}f'^2 + \frac{3}{16} a B^2 \tr^2 e^{-\c\phi}\phi'^4
-\tr^2 \tilde{\Lambda}e^{\lambda\phi} = 0\,,~~
\label{fe1}
\ena
\bea
\delta'(D-2)\tr h + \frac12 \tr^2 {\phi'}^2
- 2(D-2)_3\, \c e^{-\c\phi}(k-B)(\phi''-\c {\phi'}^2)
-\frac{3}{8} a \tr^2 B \phi'^4 e^{-\c\phi}=0 \,,
\label{fe2}
\ena
\bea
(e^{-\d} \tr^{D-2} B \phi')' &=& \c (D-2)_3 e^{-\c\phi-\d} \tr^{D-4}
\Big[ (D-4)_5 \frac{(k-B)^2}{\tr^2} + 2(B'-2\d' B)B' -4(k-B)BU(r) \nn
&& -4\frac{D-4}{\tr}(B'-\d'B)(k-B) \Big]
+\c \tr^{D-2} e^{-\c\phi} \Big(\frac{1}{2}e^{\d}f'^2
+ \frac{3}{16} a e^{-\d} B^2\phi'^4\Big) \nn
&& + \frac{3}{4} a \tr^{D-2} B\phi'^2 e^{-\d-\c\phi}\Big[3B\phi''
+\Big\{2B'-\Big(\d'+\c\phi'
-\frac{D-2}{r}\Big)B\Big\}\phi' \Big]
+ \tr^{D-2}\lambda \tilde{\Lambda}e^{-\delta+\lambda\phi},
\label{fe3}
\ena
\bea
(f'e^{\d-\c\phi}\tr^{D-2})'=0,
\label{fe4}
\ena
where we have defined the dimensionless variables: $\tr = r/\sqrt{\a_2}$,
$\tilde \Lambda = \a_2 \Lambda$, and the primes in the field equations
denote the derivatives with respect to $\tr$. Namely we measure our length
in the unit of $\sqrt{\a_2}$.
In what follows, we omit tilde on the variables for simplicity.
We have also defined
\bea
(D-m)_n &=& (D-m)(D-m-1)(D-m-2)\cdots(D-n), \nn
\label{h-def}
h &=& 1+2(D-3) e^{-\c\phi} \Big[ (D-4) \frac{k-B}{r^2}
 + \c \phi'\frac{3B-k}{r}\Big], \\
\label{tilh-def}
\tilde h &=& 1+2(D-3) e^{-\c\phi} \Big[(D-4)\frac{k-B}{r^2}
+\c\phi'\frac{2B}{r} \Big],
\ena
\vs{-5}
\bea
U(r) &=& (2 \tilde h)^{-1} \Bigg[ (D-3)_4 \frac{k-B}{r^2 B}
-2\frac{D-3}{r}\Big(\frac{B'}{B}-\d'\Big) -\frac12 \phi'^2 \nn
&& \hs{10} + (D-3)e^{-\c\phi} \Bigg\{ (D-4)_6 \frac{(k-B)^2}{r^4 B}
- 4 (D-4)_5 \frac{k-B}{r^3}\Big(\frac{B'}{B}-\d'-\c\phi'\Big) \nn
&& \hs{10} -4(D-4)\c \frac{k-B}{r^2}\Big( \c \phi'^2 +\frac{D-2}{r}\phi'-\Phi \Big)
+8 \frac{\c\phi'}{r} \biggl[\Big(\frac{B'}{2}-\d' B\Big)\Big(\c\phi'-\d'
+\frac{2}{r} \Big) \nn
&& \hs{20} -\frac{D-4}{2r}B' \biggr] + 4(D-4)\Big(\frac{B'}{2B}-\d' \Big)
\frac{B'}{r^2}-\frac{4\c}{r}\Phi (B'-2\d'B)\Bigg\} +\frac{1}{2B}e^{2\d-\c\phi} f'^2 \nn
&& \hs{20} +\frac{3}{16} a B \phi'^4 e^{-\c\phi}
-\frac{1}{B} {\Lambda}e^{\lambda\phi}\Biggr],
\\
\Phi &=& \phi'' +\Big(\frac{B'}{B}-\d' +\frac{D-2}{r}\Big) \phi'.
\label{dil}
\ena

The field equation for the Maxwell field~\p{fe4} is easily integrated to give
\bea
f'=\frac{q}{r^{D-2}} e^{\c\phi-\d},
\ena
where $q$ is a constant corresponding to the charge.
It should be noted that the quantity $q$ in this equation is actually the rescaled
one $\tilde{q}=q/ \alpha^{(D-3)/2}$ related to the real charge $q$. But as we mentioned
before, tilde is omitted here and below.

Our task is reduced to setting boundary conditions for the metric functions $B,\: \d$ and
the dilaton field $\phi$ and integrate the above set of equations, just like in our previous
papers~\cite{GOT1,GOT2,OT3,OT4,OT5}.

%%%%%%%%%%%%%%%%%%%%%%%%%%%%%%%%%%%%
%%%%%%%%%%%%%%%%%%%%%%%%%%%%%%%%%%%%
\subsection{Boundary conditions and asymptotic behaviors}
%%%%%%%%%%%%%%%%%%%%%%%%%%%%%%%%%%%%
%%%%%%%%%%%%%%%%%%%%%%%%%%%%%%%%%%%%

In this paper we consider non-extreme solutions which has a non-degenerate black hole horizon.
Hence at the horizon $r_H$, %we should have
\bea
B_H=0,~~
B_H' \neq 0,
\ena
where $B_H=B(r_H)$.
%This means that the horizon should not be degenerate.
Here and in what follows, quantities evaluated at the horizon are represented by a subscript $H$.
At the horizon, Eq.~\p{fe1} gives
\bea
h_H B_H' & =&\frac{(D-3)k}{r_H}+k^2 (D-3)_5\frac{e^{-\c\phi_H}}{r_H^3}
-\frac{q^2}{2(D-2)}\frac{e^{\c\phi_H}}{r_H^{2D-5}}
-\frac{r_H \Lambda}{D-2}e^{\la\phi_H}, \nn
& =& \frac{(D-3)k}{r_H}+k^2 (D-4)_5\frac{C}{2r_H}
-\frac{D-3}{D-2}\frac{q^2}{r_H^{2D-3}C}-\frac{r_H \Lambda}{D-2}e^{\la\phi_H},
\label{B_H}
\ena
where %we have defined
\bea
C=\frac{2(D-3)e^{-\c\phi_H}}{r_H^2}.
\ena
Henceforth we restrict our consideration to $k=1$ and $\Lambda=0$ case.
Combining Eq.~\p{B_H} with Eq.~\p{fe3} evaluated at the horizon, we obtain the quadratic equation
determining $\phi_H'$:
%\bea
%&& C \c r_H^{2 D+2} \Big[ -2(D-3) \Big\{C \Big((D-2) \c^2 [2 C (D-3)-1]
%+D-4\Big)+1\Big\} q^2 r_H^4 \nn
%&& +C (D-2)r_H^{2 D} \Big\{C (D-4) \Big(C (D-2) \c^2 [2 C (D-5) (D-3)+3 D-11]+C (D-4)_5
%+3D-11\Big)+2 (D-3)\Big\}\Big] \phi_H'^2 \nn
%&& + \Big[-C (D-2) r_H^{4 D+1} \Big\{C^2 (D-4)(D-1)_2 \gamma^2 [(D-4)_5 C^2 -2C-2]
%+[C(D-4)_5+2 (D-3)] [C (D-4)+1]^2\Big\} \nn
%&& +2 (D-3)q^2 r_H^{2 D+5} \Big\{ 2 C(C-1)(D-2)  \gamma^2 [2 C (D-4)+1]
%+[C (D-4)+1]^2\Big\}+4C(D-3)^2 q^4 r_H^9 \c^2 \Big]\phi_H' \nn
%&& -\frac{1}{2} C^2 (D-2)^2 (D-1) \gamma \Big\{C (D-4) [C (D-4) (D+1)+4]-2(D-2)\Big\}
%r_H^{4 D}
%+2 (D-3)^2 q^4 r_H^8 \c
%\nn
%&& -2 (D-2)_3 q^2 \c [3(D-4)C^2+6C-1] r_H^{2 D+4}
%=0.
%\label{iniphi}
%\ena
%
\bea
A_2 \phi_H'^2 + A_1 \phi_H' +A_0 =0
\label{iniphi}
\ena
where
\bea
A_2 &=& C \c r_H^{2 D+2} \Big[ -2(D-3) \Big\{C \Big((D-2) \c^2 [2 C (D-3)-1]
+D-4\Big)+1\Big\} q^2 r_H^4 \nn
&& +C (D-2)r_H^{2 D} \Big\{C (D-4) \Big(C (D-2) \c^2 [2 C (D-5) (D-3)+3 D-11]+C (D-4)_5
+3D-11\Big)+2 (D-3)\Big\}\Big] \nonumber
\\
&&
\\
A_1 &=&  -C (D-2) r_H^{4 D+1} \Big\{C^2 (D-4)(D-1)_2 \gamma^2 [(D-4)_5 C^2 -2C-2]
+[C(D-4)_5+2 (D-3)] [C (D-4)+1]^2\Big\} \nn
&& +2 (D-3)q^2 r_H^{2 D+5} \Big\{ 2 C(C-1)(D-2)  \gamma^2 [2 C (D-4)+1]
+[C (D-4)+1]^2\Big\}+4C(D-3)^2 q^4 r_H^9 \c^2
\\
A_0 &=&  -\frac{1}{2} C^2 (D-2)^2 (D-1) \gamma \Big\{C (D-4) [C (D-4) (D+1)+4]-2(D-2)\Big\}
r_H^{4 D}
+2 (D-3)^2 q^4 r_H^8 \c
\nn
&& -2 (D-2)_3 q^2 \c [3(D-4)C^2+6C-1] r_H^{2 D+4}
\ena
The $(\partial \phi)^4$ term in Eq.~(\ref{action1}) does not contribute to
the boundary condition of the dilaton field at the horizon.

In the asymptotic region of $r\to \infty$, we assume
\bea
B &\to& 1-\frac{2M}{r^{D-3}}+\cdots
\\
\delta &\to& 0
\label{bdcaf_2}
\\
\phi &\to& 0
\label{bdcaf_3}
\ena
where $M$ is a constant corresponding to the mass of the black hole.
Although $\delta \to \delta_{\infty}$ and $\phi \to \phi_{\infty}$, where $\delta_{\infty}$ and
$\phi_{\infty}$ are constant, these constants can be rescaled out to satisfy
Eqs.~\p{bdcaf_2} and \p{bdcaf_3} by the symmetries shown below,
and the solutions are asymptotically flat.

%%%%%%%%%%%%%%%%%%%%%%%%%%%%%%%%%%%%
\subsection{Symmetry and scaling}
%%%%%%%%%%%%%%%%%%%%%%%%%%%%%%%%%%%%

It is useful to consider several symmetries of our field equations
(or our model).

First, our field equations~\p{fe1} -- \p{fe3} have a shift symmetry:
\bea
\phi \to \phi-\phi_{\ast}, ~~
r \to e^{\gamma\phi_{\ast}/2}r, ~~
q \to e^{(D-2)\c\phi_{\ast}/2} q, ~~
(\Lambda \to e^{(\la-\c)\phi_{\ast}} \Lambda),
\label{sym2}
\ena
where $\phi_{\ast}$ is an arbitrary constant.\footnote{
There are typos in our previous papers~\cite{OT3,OT4} in the exponent of
the transformation rule in the second term; they should have the $+$ sign.}
%%%%%%
This changes the magnitude of the cosmological constant when we consider black hole
solutions in its presence. Hence this may
be used to generate solutions for different values of the cosmological constant,
given a solution with some value of cosmological constant.
Even without the cosmological constant, this symmetry can be used to change
the asymptotic value of the dilaton field.

The second one is another shift symmetry under
\bea
\delta \to \delta - \delta_{\ast}, ~~
t \to  e^{-\delta_{\ast}}t,
\label{sym3}
\ena
with an arbitrary constant $\delta_{\ast}$, which may be used to shift
the asymptotic value of $\delta$ to zero.\footnote{
For $k=0$, there is another symmetry. The field equations~\p{fe1}--\p{fe3} are
invariant under the scaling transformation
$
B \to c^2 B,
r \to c r,
$
with an arbitrary constant $c$.
If a black hole solution with the horizon radius $r_H$ is obtained,
we can generate solutions with different horizon radii but the same $\Lambda$
by this scaling transformation.}
%%%%%%

%%%%%%%%%%%%%%%%%%%%%%%%%%%%%%%%%%%%
%%%%%%%%%%%%%%%%%%%%%%%%%%%%%%%%%%%%
\section{Thermodynamical Variables}
\label{sec3}
%%%%%%%%%%%%%%%%%%%%%%%%%%%%%%%%%%%%
%%%%%%%%%%%%%%%%%%%%%%%%%%%%%%%%%%%%

Here we briefly summarize thermodynamical quantities of black holes to be
used in the following discussions.
The Hawking temperature is given by the periodicity of the Euclidean time
on the horizon as (keeping $k$ arbitrary)
\bea %----------------
\label{temp}
T_H \!\!\!\!\! && =\frac{e^{-\d_H}}{4\pi}B_H'
\nonumber \\
&& =\frac{e^{-\d_H}}{4\pi h_H}
\biggl[\frac{(D-3)k}{r_H} +\frac{(D-3)_5k^2}{r_H^3} e^{-\c\phi_H}
-\frac{q^2}{2(D-2)}\frac{e^{\c\phi_H}}{r_H^{2D-5}}
\biggr].
\label{temperature}
\ena %----------------

Along the definition of entropy in Ref.~\cite{Wald}, which originates from
the Noether charge associated with the diffeomorphism invariance of
the system, we obtain
\bea %----------------
S=-2\pi \int_\Sigma \frac{\pa {\cal L}}{\pa R_{\mu\nu\rho\sigma}}
\e_{\mu\nu} \e_{\rho\sigma},
\ena %----------------
where $\Sigma$ is the event horizon $(D-2)$-surface, ${\cal L}$ is the Lagrangian
density, $\e_{\mu\nu}$ denotes the volume element binormal to $\Sigma$.
This entropy has desirable properties such that it obeys the first law of black
hole thermodynamics and that it is expected to obey even the second law~\cite{Jacobson}.
For our present model, this gives
\bea %----------------
S = \frac{r_H^{D-2}\Sigma_{k}}{4}
\left[1+2(D-2)_3 \frac{ke^{-\c\phi_H}}{r_H^2} \right]- S_{\rm min}.
\label{entropy}
\ena %----------------
It is noted again that  $\Sigma_k$ is the volume of the unit constant curvature space, and
$\Sigma_1= \frac{2 \pi^{(D-1)/2}}{\Gamma((D-1)/2)}$.
$S_{\rm min}$ is added to make the entropy non-negative~\cite{Clunan}.

%%%%%%%%%%%%%%%%%%%%%%%%%%%%%%%%%%%%
%%%%%%%%%%%%%%%%%%%%%%%%%%%%%%%%%%%%
\section{Non-dilatonic black hole solution}
\label{sec4}
%%%%%%%%%%%%%%%%%%%%%%%%%%%%%%%%%%%%
%%%%%%%%%%%%%%%%%%%%%%%%%%%%%%%%%%%%

It will be instructive to compare our results with the non-dilatonic case.
%Let us first derive physical quantities for this case.
When the dilaton field is absent (i.e., Einstein-Maxwell-GB system), we substitute $\phi\equiv 0$
and $\gamma=0$ into Eqs.~\p{fe1} and \p{fe2}. In the $D=4$ case, the GB
term is total divergence and does not give any contribution to the field
equations. As a result, the solution reduces to the Reissner-Nordstr\"om (RN) solution.

For $D\geq 5$, the field equations can be integrated to yield~\cite{TM}
\bea %----------------
\bar{B}=1-\frac{2\bar{m}}{r^{D-3}},
\label{ND-B}
\ena %----------------
\bea %----------------
\delta = 0,
\ena %----------------
where
\bea %----------------
&&\bar{m}=\frac{r^{D-1}}{4(D-3)_4}
\Biggl[-1 \pm \sqrt{1+\frac{8(D-3)_4\bar{M}}{r^{D-1}}-\frac{(D-4)q^2}{8(D-2)r^{2(D-2)}}}
 \Biggr],
\label{ND-M}
\ena %----------------
and $\bar{M}$ is an integration constant corresponding to the asymptotic value
$\bar{m} (\infty)$ for the plus sign in Eq.~\p{ND-M}.
In the $\alpha_2\to 0$ limit, the solutions with the plus sign approach
the RN solutions. This means that they can be considered to be
the solution with GB correction to general relativity (GR). On the other hand,
the solutions with the minus sign do not have such a limit.
For these reasons, we call the solutions with plus (minus) sign the (non-)GR branch.

For $\bar{M}=0$ and   $q=0$, the metric function becomes
\bea %----------------
\label{metric-f1}
\bar{B}=\left\{
\begin{array}{ll}
1& (\mbox{GR branch}) \\
1+\dfrac{r^2}{{\ell}_{\rm eff}^2}  & (\mbox{non-GR branch})
\end{array}
\right.
\ena %----------------
where ${\ell}_{\rm eff}^2=(D-3)_4$. Hence the spacetime is Minkowski
in the GR branch while the spacetime is anti-de Sitter in the non-GR
branch although the cosmological constant $\Lambda$ is absent.

For the charged solution, besides the central singularity at $r=0$, there can be
another known as the branch singularity at finite radius $r_b>0$,
which is obtained by the condition that the inside of the square root
in Eq.~(\ref{ND-M}) vanishes. We find the $\bar{M}$-$r_b$ relation
\begin{equation} %----------------
\label{branch-sing}
\bar{M}=\frac{r_b^{D-1}}{8(D-3)_4}\biggl[ \frac{(D-4)q^2}{8(D-2)r_b^{2(D-2)}} -1\biggr].
\end{equation}   %----------------
This implies that the branch singularity can appear for positive mass in charged black hole,
but only for negative mass parameter in the neutral black hole.

It can be shown that there is no black hole solution in the non-GR branch.
On the other hand, in the GR branch, Eq.~\p{ND-M} evaluated at the horizon
$\bar B=0$ gives
\begin{equation} %----------------
\bar{M}=\frac12 r_H^{D-5} \biggl[ r_H^2 +(D-3)_4 + \frac{q^2}{8(D-2)_3 r_H^{2D}}\biggr].
\label{nondil}
\end{equation} %----------------
This is the $\bar{M}$-$r_H$ relation for the black hole without the dilaton field.
For $q=0$, we see that $D=5$ is special because only in this case the mass goes
to a finite value in the limit of $r_H \to 0$, whereas it vanishes for other dimensions
including four. This limit does not exist for charged ones ($q \neq 0$).

%%%%%%%%%%%%%%%%%%%%%%%%%%%%%%%%%%%%
%%%%%%%%%%%%%%%%%%%%%%%%%%%%%%%%%%%%
\section{$D=4$ Black Hole}
\label{secd=4}
%%%%%%%%%%%%%%%%%%%%%%%%%%%%%%%%%%%%
%%%%%%%%%%%%%%%%%%%%%%%%%%%%%%%%%%%%

For $D=4$, Eq.~\p{iniphi} to determine $\phi_H'$ reduces to
\bea
Cr_H^6 \gamma \Big\{2 C r_H^4- q^2 \big[2 C (2 C-1) \gamma ^2+1\big]\Big\} \phi_H'^2
+ r_H\Big\{ 2 C q^4 \gamma ^2+ q^2 r_H^4 \big[4C (C-1) \gamma ^2+1 \big]-2 C r_H^8\Big\}
 \phi_H' \nn
+ \gamma  \big[12 C^2 r_H^8-2 (6 C-1) q^2 r_H^4+ q^4\big]
=0.
\label{p4}
\ena
The discriminant for $\c=\frac12$ is
\bea
\frac{1}{4} r_H^2 (q^2-2 C r_H^4)^2 \big[C^2 q^4+2 C (6 C^2-3
   C+4) q^2 r_H^4+4 (1-6 C^2) r_H^8\big].
\label{discre_d4}
\ena

Our procedure for obtaining solutions is as follows.
First we choose the suitable values of  parameter $q$ and the boundary condition $\phi_H$
(say $\phi_H=0$) at the horizon.
Given the horizon radius $r_H$ together with $\phi_H'$ determined
by \p{p4}, we integrate the field equations~~\p{fe1} -- \p{fe3} outward from
the horizon numerically. There are two possible solutions for $\phi_H'$ in Eq.~\p{p4},
but only the smaller one  gives the black hole solutions.
Although the dilaton field takes nonvanishing value $\phi_{\infty}$ in the asymptotic
region generically, it can be set to zero by using the shift symmetry~\p{sym2}.
Note that the variables defined by
\bea
x = e^{\gamma\phi_H/2}r_H, \quad
y = e^{(D-2)\gamma\phi_H/2} q,
\label{var}
\ena
do not change under this shift symmetry.
Then we redo this procedure by changing $r_H$ but keeping the values of charge $q$ and $\phi_H$
unchanged, which means that  $y$ is also unchanged, and obtain another
solution with the different asymptotic value of the dilaton field $\phi_{\infty}$.
Using the shift symmetry to set  $\phi_{\infty}$ to zero, we have the solution
with the same $y$ but different $x$.
In this way we obtain various solutions for one chosen $y$.
Next, we repeat this procedure for different $y$ (practically, with different $q$ and
the same $\phi_H$).
After this process, we not only obtain solutions but also identify the parameter ranges
where the solutions exist on the $x$-$y$ plane.

Since the values of $x$ and $y$
are unchanged by the shift symmetry~\p{sym2},
it is convenient to draw various diagrams in terms of these variables.
$x$ and $y$ will be referred to as the ``scaled horizon radius'' and the ``scaled charge'',
respectively.
After identifying the  parameter region where solutions exist, we can change
these ``scaled variables" to physical ones.

%--figures-------------------------------------------------------------
\begin{figure}[htb]
\begin{center}
\includegraphics[width=16cm]{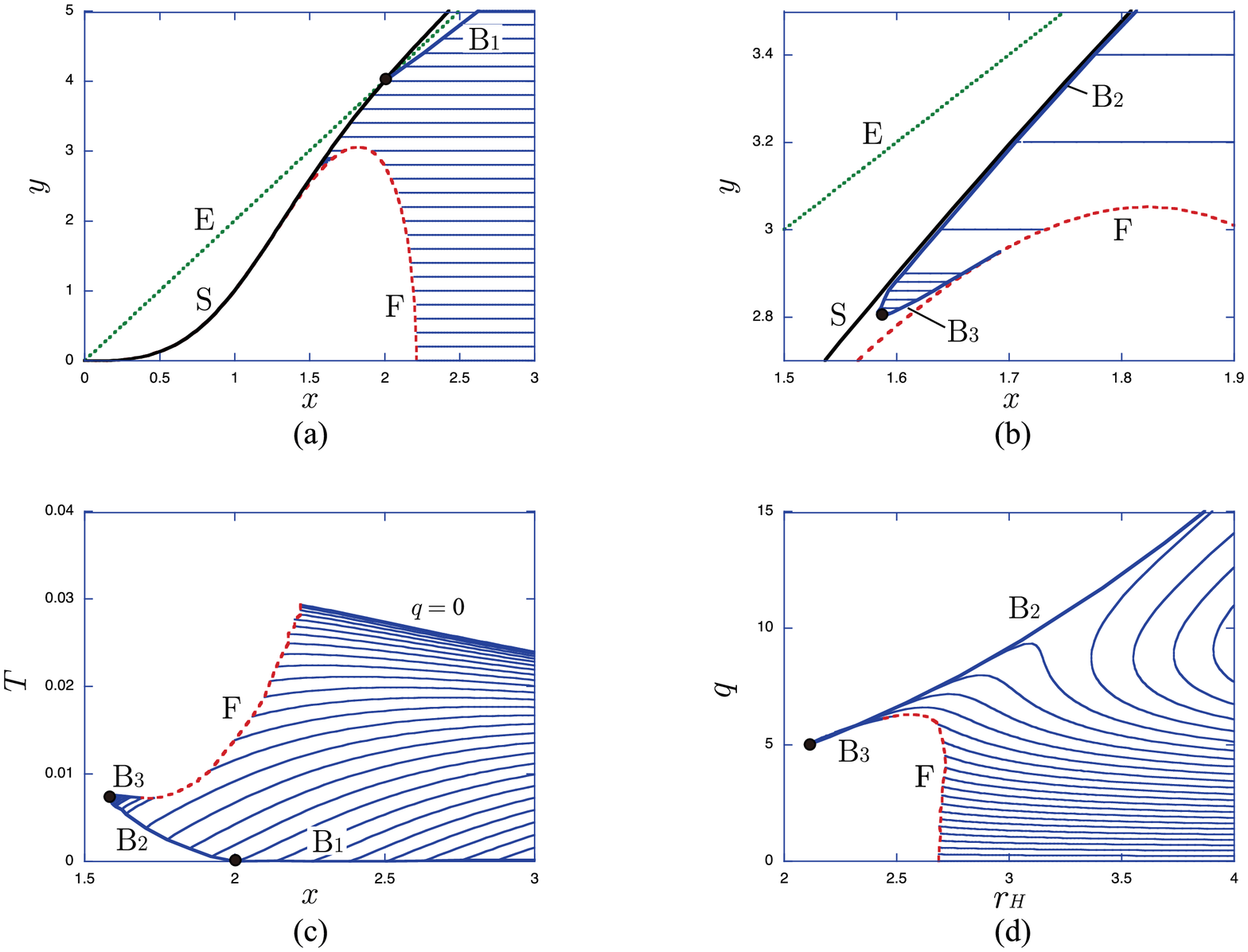}
\end{center}
\vspace*{-10mm}
\caption{
The parameter region where the black hole solution exists in $D=4$.
The black hole solution exists in the allowed region which is
shadowed by (blue) thin lines.
(a) The horizontal axis is $x = e^{\gamma\phi_H/2}r_H$ and the vertical axis
is $y = e^{\gamma\phi_H} q$.
On the dashed (red) curve F, the discriminant~(\ref{discre_d4}) is zero.
Below this line, the discriminant is negative and there is no appropriate boundary
condition on the event horizon. On the curve S the first derivative of the dilaton field
on the horizon $\phi_H'$ diverges and the solution becomes singular.
The line E represents the extreme solution. On the curve B$_1$, $\phi_{\infty}$
diverges numerically which means $\phi_H$ diverges physically by the shift symmetry.
(b) The magnified diagram around the top of the forbidden region.
Around the curve B$_2$ numerical calculation becomes
unstable and we cannot find the black hole solution above it.
On the curve B$_3$, the third derivative of the dilaton field diverges
while the Kretschmann invariant does not.
(c) The allowed region in $x$ and the physical temperature $T$ of the black hole.
The temperature becomes zero on the line B$_1$. The reason is that the horizon of the
solutions on it have infinite size.  The solutions on the boundaries F, B$_2$, and B$_3$
have non-zero finite temperature.
(d) The diagram of the allowed region in terms of the physical quantity $r_H$ and $q$.
For each value of the charge $q$, there is a lower bound for the horizon radius $r_H$.
}
\label{d4-1}
\end{figure}
%--figures-------------------------------------------------------------

We find that there are black hole solutions in the region shadowed by horizontal
thin (blue) lines in Fig.~\ref{d4-1}(a).
The horizontal and vertical axes are $x$ and $y$, respectively.
These horizontal (blue) lines actually consist of the sequences of solutions obtained
by the above procedure.
Although the lines are drawn for separate distance in $y$ with intervals $\Delta y =0.2$,
it should be understood that there are solutions in the whole shadowed region.
We call this region (where the black hole solutions exist) {\it the allowed region}.
On the dashed (red) curve F determined by
\bea
\mbox{F~:}~~~ y^2 = \left[- 2x^4 +3x^2-12 + \sqrt{3(x^8-4x^6+27 x^4-24 x^2+48)} \right] x^2,
\ena
the discriminant~(\ref{discre_d4}) vanishes.
Below the curve F, the discriminant is negative and the values of $\phi_H'$ becomes imaginary,
and hence there is no black hole solution. Furthermore, on the curve F, it can be found
analytically that the second derivative of the dilaton field diverges at the horizon.
The Kretschmann invariant
\begin{eqnarray}
{\cal K}=&& \!\!\!\!\!
R^{\mu\nu\rho\sigma}R_{\mu\nu\rho\sigma}
\nonumber \\
= && \!\!\!\!\!
\bigl[B''-3B'\d'+2B(\d'^2-\d'')\bigr]^2 +\frac{2(D-2)}{r^2} (B'^2-2BB'\d'+2B^2 \d'^2)
+ \frac{2(D-2)_3}{r^4} (k-B)^2,
\end{eqnarray}
also diverges there,
and the solution is singular.
We call the region on and below the curve F {\it the forbidden region}.
For $q=0$, the boundary is $x=24^{1/4}\approx 2.213$.
For $y>3.0523$ (above the top of the curve F), the discriminant is positive for all
$x$ (and  $r_H$).

The extremal condition given by $T_H=0$ (or $B'_H=0$)  is written in general dimensions as
\bea
\mbox{E~:~~~} y^2 = 2\bigl[(D-2)_3 x^{2(D-3)}+(D-2)_5 x^{2(D-4)}\bigr].
\ena
In four dimensions, this reduces to
\bea
\mbox{E~:~~~} y=2x,
\ena
and is depicted by the dotted (green) line E.
Above the line E, $B_H'$ is negative and the horizon $r_H$ is not a black hole horizon
(i.e., a closed trapped surface).
Actually we find that there is no black hole solution beyond
the solid (black) curve S, on which $\phi'_H$ diverges.
Hence the curve S is expected to give a boundary of the allowed region (for $y<4$).
The equation of S can be read off from Eq.~\p{p4},
as the condition that the coefficient of $\phi'^2_H$ vanishes:
\bea
\mbox{S~:~~~} y=2x^3\sqrt{\frac{1}{x^4-x^2+4}}.
\ena
For small $x\;(\lsim 1)$, this singular curve S almost overlaps with the curve F.
And it crosses with the extreme curve E at $(x,y)=(2,4)$.

For $y\geq 4$, we find that the black hole solutions exist to the right of the boundary
B$_1$. On B$_1$, we find that the asymptotic value of the dilaton field $\phi_{\infty}$
diverges in the numerical integration, which means that $\phi_H$ diverges after $\phi_H$
is shifted such that $\phi_{\infty}\to 0$ by the shift symmetry.
Figure~{\ref{d4-1}}(b) is a magnified diagram around the top of forbidden region.
For $3.0523<y<4$, we find that there seems to be a boundary B$_2$ slightly to
the right of the curve S.
We have confirmed that the black hole solutions exist on the right side of B$_2$.
In our numerical analysis, the calculation becomes unstable and stops for parameters
just out of the event horizon before we let the parameters reach the curve S,
which strongly suggests that there exists the boundary B$_2$.
On the other hand, expanding the field equations around $r_H$,
we do not find any singular behavior in the right region of S including B$_2$ analytically.
Unfortunately within our numerical accuracy, it is difficult to determine precisely
where the true boundary is. However, there is certainly the boundary B$_2$ different from S
around $y \sim 2.8$, and it is plausible that it merges with S at the point $(x,\:y)=(2,\:4)$.

For $y<3.0523$, the allowed region splits into two parts.
In the region to the right of the curve F, the black hole solutions exist.
On the left side of the curve F, we also have what we call {\it solitary solutions}.
We find that the right boundary of the region where the solitary solutions exist
is partly F and partly B$_3$ (see Fig.~1(b)), where the third derivative of the dilaton field
diverges at certain radius $r>r_H$ while the Kretschmann invariant ${\cal K}$ does not.
The endpoint of B$_2$ and B$_3$ is depicted by a (black) dot.
Below this, there are black hole solutions only to the right of F.

In Fig.~\ref{d4-1}(c), we show the temperature vs. the scaled horizon radius relation.
There are solutions in the region where (blue) thin curves are plotted.
It should be noted that the temperature vanishes on the boundary B$_1$,
although B$_1$ is different from the extreme curve E.
The reason why the temperature vanishes is that the asymptotic value
of the dilaton field $\phi_{\infty}$ gets infinite and the horizon radius $r_H$ and the charge
$q$ of the black hole  on B$_1$ become infinitely large after using the shift symmetry.
The physical value of $\phi_H$ diverges and the solutions on B$_1$ is singular.
On the left lower boundary B$_2$ around $x=1.6 \sim 2$
the temperature is non-zero.
The right-upper boundary is determined by $q=0$, beyond which there is no solution.

%--figures-------------------------------------------------------------
\begin{figure}
\vspace*{-5mm}
\begin{center}
\includegraphics[width=16cm]{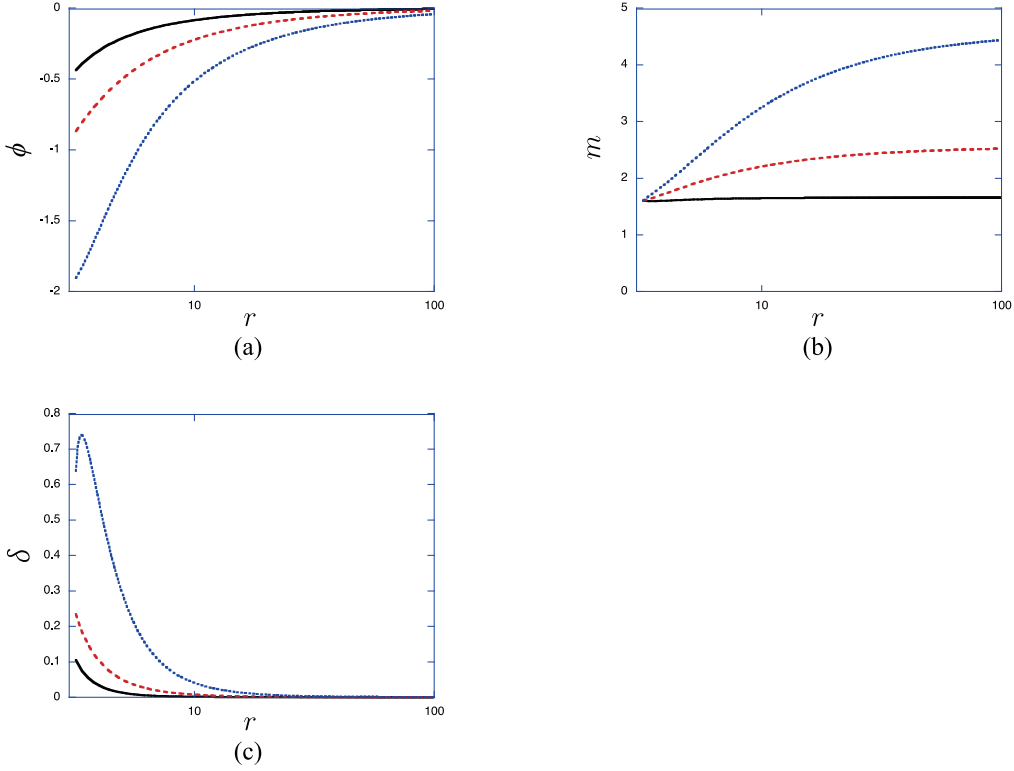}
\end{center}
\vspace*{-9mm}
\caption{Configurations of (a) the dilaton field $\phi$,  (b) the mass function $m$, and
(c) the lapse function $\delta$ of
the black hole solutions in $D=4$.
The horizon radii are $r_H=3.2$, and the charges are $q=0$ (solid (black) line),
 $q=5$ (dashed (red) line),  $q=10$ (dotted (blue) line).
}
\label{d4-config}
\end{figure}
%--figures-------------------------------------------------------------

The diagrams drawn in terms of $x$ and $y$ are useful to identify the parameter
region numerically where the black hole solutions exist. To discuss the physical
properties, however, it is more convenient to show the diagrams in terms of
the physical quantities such as  $r_H$ and $q$ directly.
In Fig.~\ref{d4-1}(d), we display the allowed region in terms of the charge $q$
and the horizon radius $r_H$. There is a lower bound on the horizon radius for
any value of charge $q$.
When the charge is small ($q\lsim 5$), the curve F gives the boundary.
We can see that the minimum size of the black hole is almost independent
of the charge in this range.
When the charge is in the range $5\lsim q \lsim 6.3$, the curve F is again
the boundary, but it is not the lower bound;  there are solitary solutions in the thin region
in the left part of the diagram. This cusp structure is surrounded by B$_2$ and B$_3$.
The structure is so thin that it is difficult to create the black hole
in this region in physical processes.
For $q\gtrsim 6.3$, the lower boundary is given by B$_2$.
The curve B$_1$ with $y\geq 2$ in Fig.~\ref{d4-1}(a) is pushed to the upper right
infinitely ($r_H\to\infty$, $q\to\infty$).

So far we have presented the discussion of the parameter region where black hole solutions
exist. We now focus on how the field functions of the black hole solutions behave for
a typical boundary condition.
Figure~\ref{d4-config} shows the behaviors of the function $\phi$, $m$ and $\d$
for the horizon radius $r_H=3.2$ for neutral and charged cases with $q=0,5$ and 10.
This horizon radius is chosen such that the solutions exist for these charges.
For $q=10$, the parameters are close to the boundary B$_2$ in the diagram Fig.~\ref{d4-1}(d).
We see that these functions have smooth behaviors.
The dilaton field $\phi$ increases monotonically for each charge (Fig.~\ref{d4-config}(a)).
The mass function $m$  of the neutral solution decreases near the horizon and increases
towards a finite value as $r$ increases~\cite{GOT1}. On the other hand the mass function
of the charged solution increases all the range from the horizon (Fig.~\ref{d4-config}(b)).
The lapse function $\delta$ decreases for small $q$. As the charge becomes large and
the parameters approach B$_2$, $\delta$ increases around the horizon and decreases to zero.
Since the dilaton field couples to the gauge field and the GB term, the effects of the dilaton
field and higher curvature terms become significant as the charge becomes large.

Up to this point, our discussions are rather qualitative with only figures of regions
and behaviors of functions.
To evaluate actually some quantities, it is necessary to have quantitative results.
In order to get some idea on what are the typical physical quantities,
here we tabulate them  for the black hole solutions with the charge $q=0,\:5$
and 10 in Table \ref{table_d4}.
\begin{table}[h]
\begin{tabular}{@{~~~}r@{~~~}|c|c|@{~~~}l@{~~~}|@{~~~}l@{~~~}|@{~~~}l@{~~~}|@{~~~}l@{~~~}|c}
\hline
$q$ & ~~~$r_H$~~~ & $M$ & ~~~~~~$\delta_H$ & ~~~~~~$\phi_H$ & ~~~~~$\phi_H'$ & ~~~~~~$T$
 & $S/\Sigma_1$ \\
\hline\hline
0 & 2.8 & ~~1.50205~~ & 0.291474  &  $-$0.650704 & 0.504050  &0.0282842  &  ~~3.34452~~ \\
\cline{2-8}
 & 3.2 &  1.65494 &  0.104926 & $-$0.439547  &  0.253039 &  0.0248377 & 3.80579 \\
\cline{2-8}
 & 4.0 & 2.02371  & 0.0319963  & $-$0.256190  & 0.110000  &  0.0198896 & 5.13666  \\
\cline{2-8}
 & 6.0 & 3.00621  & 0.00521331  & $-$0.106409  & 0.0294481  & 0.0132626  &  10.0546 \\
\cline{2-8}
 & 8.0 &  4.00251 & 0.00156277  & $-$0.0586656  & 0.0120864  & 0.00994713  & 17.0298  \\
\hline
5 & 2.8 & 2.48068  &  0.502017 &  $-$1.10819 & 0.564272  &  0.0143593 & 3.70037  \\
\cline{2-8}
 & 3.2 & 2.55368  & 0.234784  &  $-$0.871222 & 0.347580  & 0.0143019  & 4.10591  \\
\cline{2-8}
 & 4.0 & 2.77132  & 0.0893922  & $-$0.582756  &  0.183430 &   0.0137251&  5.33827 \\
\cline{2-8}
 & 6.0 & 3.51862  &  0.0168288 &  $-$0.269283 &  0.0562110  & 0.0111825   & 10.1441  \\
\cline{2-8}
 & 8.0 & 4.38982  & 0.00522316  & $-$0.153149  & 0.0238626  & 0.00902933  & 17.0796  \\
\hline
10 & 2.8 & ---  & ~~~~~---  & ~~~~~~---  & ~~~~~---  & ~~~~~---  & ---  \\
\cline{2-8}
 & 3.2 &  4.57779 & 0.635740  &  $-$1.90468  & 0.285430  & 0.00099367  &  5.15176 \\
\cline{2-8}
 & 4.0 &  4.61313 & 0.529248  & $-$1.51849  &  0.413647 & 0.00404222  &  6.13667 \\
\cline{2-8}
 & 6.0 &  4.92428 & 0.112153  & $-$0.774269  & 0.154977  &  0.00651324 &  10.4728 \\
\cline{2-8}
 & 8.0 &  5.49493 & 0.0339071  & $-$0.445769  &  0.0650895 & 0.00667781  & 17.2497  \\
\hline
\end{tabular}
\caption{Typical values of the physical quantities of the black hole solutions
in $D=4$.
}
\label{table_d4}
\end{table}
%--figures-------------------------------------------------------------
\begin{figure}[h]
\vspace*{-5mm}
\begin{center}
\includegraphics[width=15cm]{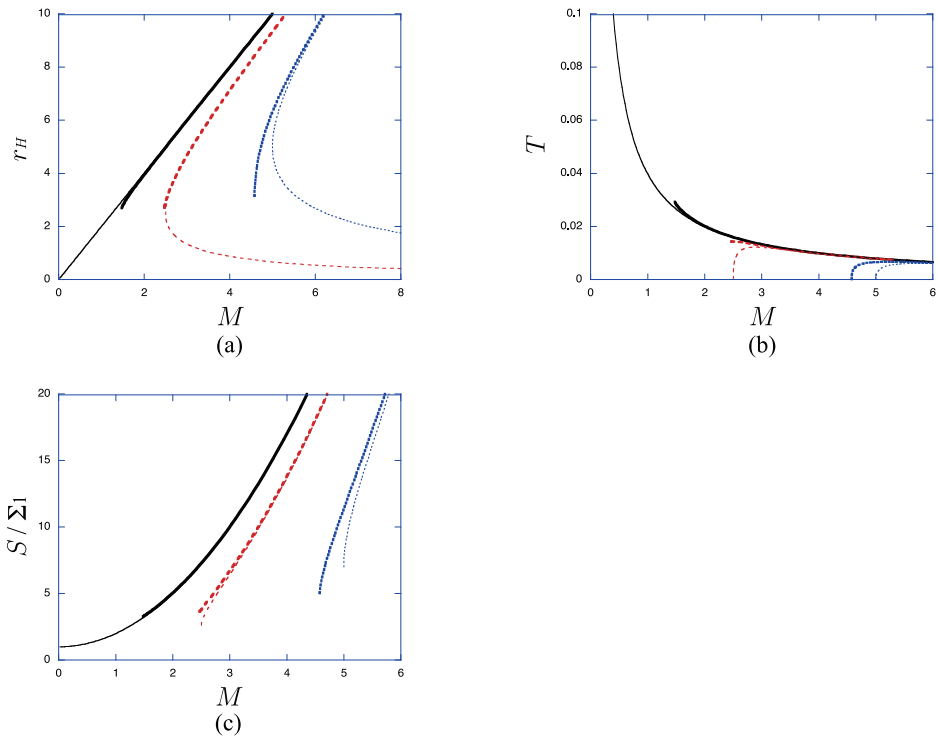}
\end{center}
\vspace*{-5mm}
\caption{Various relations for $D=4$ black hole solutions in (dilatonic) EGB systems.
(a) $M$-$r_H$ diagram,
(b) $M$-$T$ diagram,
(c) $M$-$S/\Sigma_1$ diagram.
Solid (black) lines for $q=0$,
dashed (red) lines for $q=5$,
and dotted (blue) lines for $q=10$.
Dilatonic solutions are given by thick lines and non-dilatonic ones by thin lines.
In $D=4$, the non-dilatonic solutions are Schwarzschild ($q=0$) and
RN black hole ($q\ne 0$) solutions.
}
\label{d4-3}
\end{figure}
%--figures-------------------------------------------------------------

Collecting all these information, we give summary of our results in the form
of the relations between physical quantities of the black holes.
Figure~\ref{d4-3} shows various relations for neutral case (solid (black) line),
charged case with $q=5$ (dashed (red) line) and with $q=10$ (dotted (blue) line).
For comparison, we also give our results by the thick (thin) lines for dilatonic
(non-dilatonic) solutions [see Eqs.~(\ref{ND-B})-(\ref{ND-M}) for the non-dilatonic case].
For the non-dilatonic solutions, the GB term becomes total divergence in $D=4$, and
the solutions are Schwarzschild and RN black hole solutions
($\bar{B}=1-2\bar{M}/r+q^2/4r^2$).\footnote{The normalization of our
charge is a factor 2 different from the conventional one.}
For the RN black hole, there is the minimum mass for each charge,
which corresponds to the extreme black hole solution.
There are two values of $r_H$ for the fixed charge $q(\neq 0)$ and mass
(see Fig.~\ref{d4-3}(a)).
The larger is the radius of the black hole horizon,
the smaller is that of the inner horizon of the same solution.
In contrast, in the dilatonic case, there are lower bounds on the horizon radii
as well as the masses of the black hole solutions for given fixed charges $q$.
They are given as
$(q,\; x,\; y)=$
(0,\; 2.2134,\; 0),
(5,\; 2.0117,\; 2.7593),
(10,\; 1.9686,\; 3.9144).
These solutions are not extremal.
The lower bounds on the solutions with $q=0$ and 5 are
determined by the forbidden region in Fig.~\ref{d4-1};
the parameters of the corresponding solutions are just on the boundary F.
The horizon radii of the smallest black holes are almost the same for this range of
charge as we have pointed out already, while their masses depend on the charge.
For $q=10$, the lower bound corresponds to the singular solution not on F but on B$_2$.
{}From these figures, we see that the bigger the charge is, the further
the curves move away from the neutral one, and the effect of the dilaton field is bigger.
This is expected because the bigger the charge is, the bigger the dilaton coupling is.

Figures~\ref{d4-3} (b) and \ref{d4-3} (c) give relations between the mass and temperature
and their entropy, respectively.
For the neutral case, the GB term has the tendency to raise the temperature
compared to the non-dilatonic solution. For the charged case, however,
the temperature is lower than the non-dilatonic solutions.
The non-dilatonic solutions have extremal limit with $T=0$
while all the dilatonic solutions have non-zero finite temperature.
This fact gives the following scenario. A dilatonic black hole loses
its mass by emitting radiation and continues evaporating until
the solution reaches the minimum mass solution and the spacetime becomes singular.
The entropy of the Schwarzschild and RN black holes in the EGB theory is different from
those in GR because of the second term in the square bracket of Eq.~\p{entropy}.
This term is proportional to $\alpha'$ and is absent in GR.
In $D=4$, the contribution from this term to $S/\Sigma_1$ is 1 (constant).
Hence, for instance, entropy of the Schwarzschild black hole in the zero mass limit
is $S/\Sigma_1=1$ in the EGB theory.

%%%%%%%%%%%%%%%%%%%%%%%%%%%%%%%%%%%%
%%%%%%%%%%%%%%%%%%%%%%%%%%%%%%%%%%%%
\section{$D=5$ Black Hole}
\label{secd=5}
%%%%%%%%%%%%%%%%%%%%%%%%%%%%%%%%%%%%
%%%%%%%%%%%%%%%%%%%%%%%%%%%%%%%%%%%%

For $D=5$,  Eq.~\p{iniphi} to determine $\phi_H'$ reduces to
\bea
&& C r_H^8 \c \Big\{ q^2 \big[3 C (4 C-1) \c^2+C+1\big]
-3 C r_H^6 (3 C^2 \gamma ^2+C+1) \Big\}  \phi_H'^2 \nn
&& ~~~~~~ +r_H \Big\{(C+1)^2 r_H^6 (3 C r_H^6-q^2)
-2 C\c^2 \big[9 C^2 (C+1) r_H^{12}+3 (C-1) (2 C+1) q^2 r_H^6+2 q^4\big]\Big\} \phi_H'\nn
&& ~~~~~~ +\gamma  \Big[9 C^2(3 C^2+2 C-3) r_H^{12}+3 (3 C^2+6 C-1) q^2 r_H^6-2 q^4\Big]=0.
\ena
%-
The discriminant is
\bea
&& \frac{1}{2} r_H^2 (q^2 -3 C r_H^6)^2 \big[18 C^6 r_H^{12}+30 C^5 r_H^{12}+C^4 r_H^6
(12 q^2+5 r_H^6)+2 C^3 r_H^6 (11 q^2-8 r_H^6) \nn
&& ~~~~~~ +C^2 (2 q^4+3 q^2 r_H^6-12 r_H^{12})+8 C r_H^6 (q^2+r_H^6)+2r_H^{12}\big].
\label{discd5}
\ena

\begin{figure}[bt]
\vspace*{-5mm}
\begin{center}
\includegraphics[width=16cm]{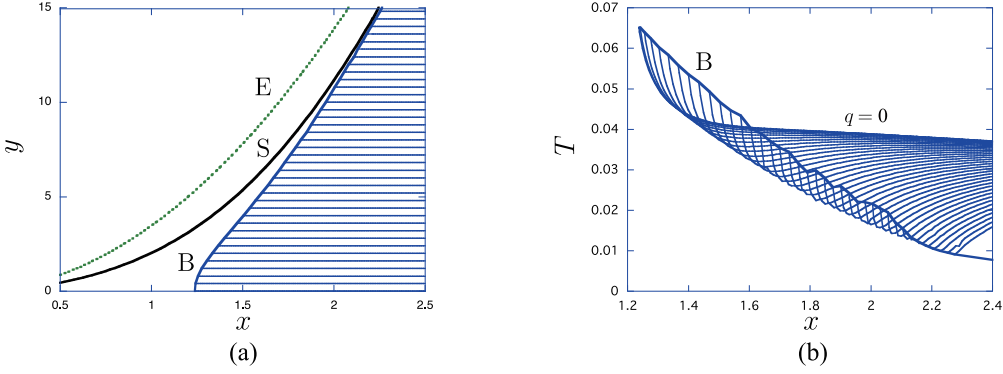}\\
\end{center}
\vspace*{-7mm}
\caption{The parameter region where the black hole solution exists in $D=5$.
The black hole solution exists in the allowed region which is
shadowed by (blue) thin lines.
(a) The horizontal axis is $x = e^{\gamma\phi_H/2}r_H$ and the vertical
axis is $y = e^{3\gamma\phi_H/2} q$.
On the curve S the first derivative of the dilaton field
on the horizon $\phi_H'$ diverges and the solution becomes singular.
The curve E represents the extreme solution.
The curve B gives the boundary where the solutions exist.
On the curve B the second derivative of the dilaton field diverges at $r>r_H$ for $y\lesssim 12$.
(b) The allowed region in terms of  $x$ and the physical temperature $T$ of the black hole.
The solutions on the boundary B have non-zero finite temperature.
}
\label{d5-3}
\end{figure}

We show the allowed region where the black hole solutions exist in Fig.~\ref{d5-3}(a).
Since the discriminant~\p{discd5} is always positive for $C>0$, there is no forbidden region.
This is also the case for higher dimensions.
There are, however, other bounds given by the curve S on which the first derivative of
the dilaton field diverges at the horizon. There is also the curve E, on which the solution
becomes extreme. The equations of these curves are
\bea
\mbox{S:~~~} y=2x^2 \sqrt{\frac{3(x^4+4 x^2+12)}{x^4+x^2+48}},
\ena
\bea
\mbox{E:~~~} y=2\sqrt{3} x^2,
\ena
respectively.
The actual boundary of the allowed region is given by the curve B.
The interval of each thin line in the allowed region is $\Delta y=0.4$. The thin lines
in other figures in this section have the same interval.
%--figures-------------------------------------------------------------
%--figures-------------------------------------------------------------
%--figures-------------------------------------------------------------
\begin{figure}[htb]
\vspace*{-5mm}
\begin{center}
\includegraphics[width=16cm]{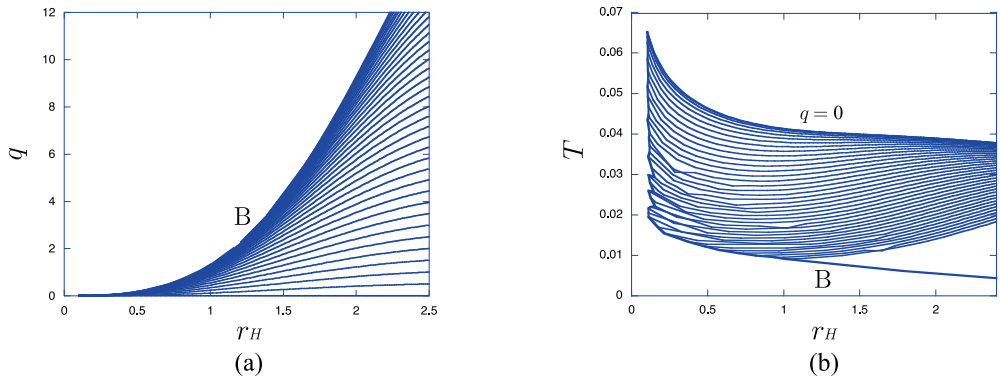}
\end{center}
\vspace*{-5mm}
\caption{The parameter region where the black hole solutions exist in $D=5$.
(a) The diagram of the horizon radius $r_H$ and the charge $q$.
For each value of the charge $q$, there is a lower bound for the horizon radius $r_H$.
(b)  The diagram of the horizon radius $r_H$ and the temperature $T$.
}
\label{d5-4}
\end{figure}
%--figures-------------------------------------------------------------

For $y\lesssim 12$, $\phi_H$ is large near the boundary B so that the physical horizon
radius $r_H$ and the charge $q$ are very small as we will see soon (recall that our physical
$\phi_H$ is determined by the shift symmetry~\p{sym2}). This implies that the solution
has the similar properties to the neutral solution. (The neural solution with higher
order term of the dilaton field ($a=1$) is summarized in the Appendix.) The neutral
solution also has the lower bound on $r_H$, where the second derivative of the dilaton
field diverges at outer region $r>r_H$. We find that the same divergence occur
also in the charged case. Kretschmann invariant  diverges there.
Hence the outer domain of the black hole with parameters on B ($y\lesssim 12$) is singular.
For larger $y~(\gtrsim12.4)$,  $\phi''$ diverges at radius close to the horizon
$r\approx r_H$ for solutions with parameters on B.
For larger $y$, the extreme curve E and S intersect at
$(x,\: y)=(2\sqrt{3}, \: 24 \sqrt{3})=(3.464, \: 41.57)$
but the boundary B is still below the extreme curve E.

In Fig.~\ref{d5-3}(b), we show the allowed region in terms of the scaled horizon radius $x$
and the temperature $T$. Each thin curve has the same $y$.
The allowed region is three-dimensional with another axis of $y$, and
the boundary looks to turn around into inside the region.
On the boundary represented by B, the temperature of the minimum size solution is
finite, hence a small black hole may evolve to this solution through the evaporating
process.
For $y\lsim12$ there may appear the singularity at non-zero radius away from the horizon.
On the other hand, for $y\gtrsim 12.4$, the singularity appears at the radius very close
to the horizon just before the radiation stops.
The qualitative difference of the boundary B with $y\lesssim12$ or $y\gtrsim12.4$
mentioned above appears around $x=2.1$ in the diagram. The boundary curve B decreases
rapidly there, and no ``fold" is observed for $x\gtrsim 2.1$.

We also give the allowed region in terms of $r_H$, $q$ and $T$ in Fig.~\ref{d5-4}.
The curve B corresponds to the boundary in  Fig.~\ref{d5-3}.
In Fig.~\ref{d5-4}(a) we see that the solutions with the same $y$ (depicted by thin lines)
converge to the lower left corner corresponding to neutral solution.
For example, the physical parameter of the solution with $(x, y)=(2.02, 11)$
is  $(r_H, q)=(0.177, 0.0848)$.
The boundary B changes its qualitative property at $(r_H, q)\sim (0.2, 0.1)$  ($y\approx 12$).
Figure~\ref{d5-4}(b) shows that the temperature gets lower in the presence of the charge
but it does not vanish for all solutions. The left vertical boundary of B corresponds to
$y\lesssim 12$.

\begin{figure}
\vspace*{-5mm}
\begin{center}
\includegraphics[width=16cm]{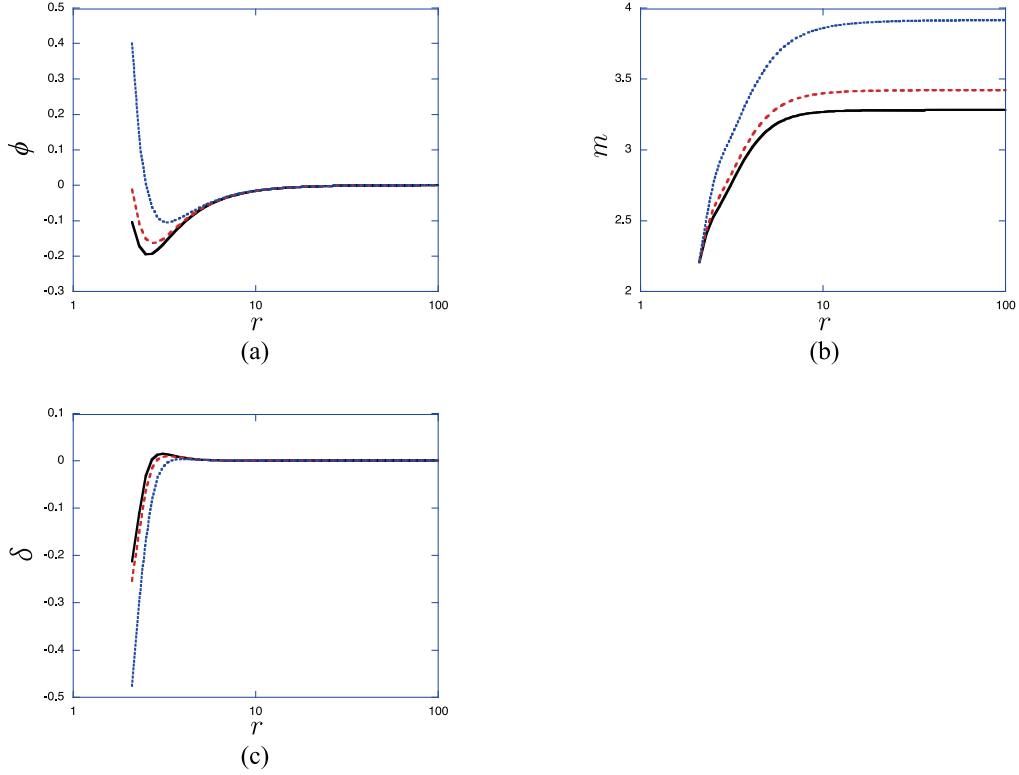}
\end{center}
\vspace*{-9mm}
\caption{
Configurations of (a) the dilaton field $\phi$,  (b) the mass function $m$, and
(c) the lapse function $\delta$ of
the black hole solutions in $D=5$.
The horizon radii are $r_H=2.1$, and the charges are $q=0$ (solid (black) line),
 $q=5$ (dashed (red) line),  $q=10$ (dotted (blue) line).
}
\label{d5-config}
\end{figure}

Figure~\ref{d5-config} shows the configurations of the field functions $\phi$, $m$ and $\d$
for the horizon radius $r_H=2.1$ for neutral and charged cases with $q=0,5$ and 10.
This horizon radius is chosen such that the solutions exist for these charges.
There are some different features in $D=5$ dimension compared to the
$D=4$ case; the dilaton field decreases and
the lapse function increases rapidly around the event horizon.
Although the configurations in the neutral case are qualitatively different depending
on the horizon radius (details can be found in Ref.~\cite{GOT1}),
we find that the variations of the field functions become large
as the charge becomes large in general due to the dilaton coupling as in the $D=4$ case.

%--figures-------------------------------------------------------------
%--figures-------------------------------------------------------------

%--figures-------------------------------------------------------------
\begin{figure}[h]
\vspace*{-10mm}
\begin{center}
\includegraphics[width=16cm]{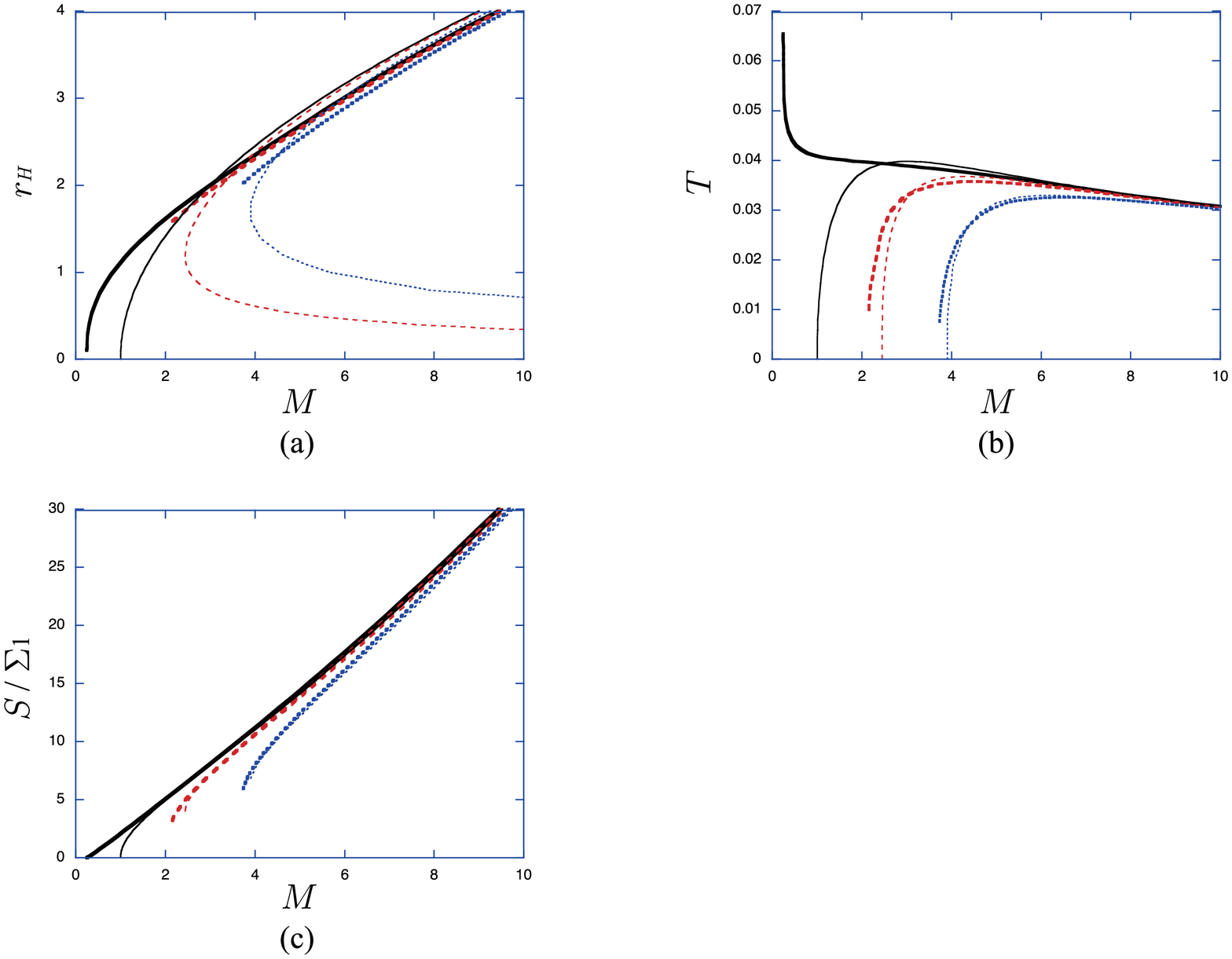}
\end{center}
\vspace*{-7mm}
\caption{
Various relations for $D=5$ black hole solutions in (dilatonic) EGB systems.
(a) $M$-$r_H$ diagram,
(b) $M$-$T$ diagram,
(c) $M$-$S/\Sigma_1$ diagram.
Solid (black) lines for $q=0$,
dashed (red) lines for $q=5$,
and dotted (blue) lines for $q=10$.
Dilatonic solutions are given by thick lines and non-dilatonic ones by thin lines.
}
\label{d5-5}
\end{figure}
%--figures-------------------------------------------------------------

%In order to gain idea on what are the typical physical quantities and have qualitative results,
Here we tabulate physical quantities for the black hole solutions with the charge
$q=0, 5$ and 10 in Table \ref{table_d5}.
\begin{table}[h]
\begin{tabular}{@{~~~}r@{~~~}|c|c|@{~~~}l@{~~~}|@{~~~}l@{~~~}|@{~~~}l@{~~~}|@{~~~}l@{~~~}|@{~~~}l@{~~~}}
\hline
$q$ & ~~~$r_H$~~~ & $M$ & ~~~~~~$\delta_H$ & ~~~~~~$\phi_H$ & ~~~~~$\phi_H'$ & ~~~~~~$T$
 & ~~~$S/\Sigma_1$ \\
\hline\hline
0 & 0.2 & ~~0.25377~~ & ~~0.512040  &  ~~7.37067 & $-$19.7397  & 0.0563593  &  0.0170506 \\
\cline{2-8}
 & 1.7 &  2.21038 &  $-$0.341768 & ~~0.268320  &  $-$1.08980 &  0.0395517 & 5.68795 \\
\cline{2-8}
 & 2.2 & 3.56697  & $-$0.173250  & $-$0.168216  & $-$0.346471  &  0.0383783 & 9.84113  \\
\cline{2-8}
 & 4.0 & 9.43189  & ~~0.065321  & $-$0.305025  & ~~0.172855  & 0.0313079  &  29.9771 \\
\cline{2-8}
 & 6.0 &  19.2458 & ~~0.0181629  & $-$0.164383  & ~~0.0680353  & 0.0237666  & 73.5419  \\
\hline
5 & 0.2 &  ---  & ~~~~~---  & ~~~~~~---  & ~~~~~---  & ~~~~~---  & ~~~~~---  \\
\cline{2-8}
 & 1.7 & 2.41969  &  $-$0.484009 & ~~0.678086  & $-$2.25502  & 0.0254686 & 4.86175   \\
\cline{2-8}
 & 2.2 & 3.69719  & $-$0.208188  &  $-$0.0987342 &   $-$0.464995&  0.0352647 & 9.59600   \\
\cline{2-8}
 & 4.0  &  9.49343 &  ~~0.064932 &  $-$0.306446  & ~~0.171797   & 0.0310818  & 29.9870 \\
\cline{2-8}
 & 6.0 & 19.2748  & ~~0.0182388  & $-$0.165188  & ~~0.0682174  & 0.0237302  & 73.5498  \\
\hline
10 & 0.2 & ---  & ~~~~~---  & ~~~~~~---  & ~~~~~---  & ~~~~~---  & ~~~~~---  \\
\cline{2-8}
 & 1.7 &   ---  & ~~~~~---  & ~~~~~~---  & ~~~~~---  & ~~~~~---  & ~~~~~---  \\
\cline{2-8}
 & 2.2 &  4.14235 & $-$0.343864  & ~~0.157902  &  $-$1.05281 & 0.0236971  &  8.76096 \\
\cline{2-8}
 & 4.0 &  9.68000 & ~~0.0637826  & $-$0.310739  & ~~0.168659  &  0.0304058 &  30.0171 \\
\cline{2-8}
 & 6.0 &  19.3627 & ~~0.0184687  & $-$0.167607  &  ~~0.068766 & 0.0236212  & 73.5735  \\
\hline
\end{tabular}
\caption{Typical values of the physical quantities of the black hole solutions
in $D=5$.
}
\label{table_d5}
\end{table}

In Fig.~\ref{d5-5}, we give the relations of physical quantities for neutral
and charged solutions with different charges as well as those in non-dilatonic theory.
For the  neutral case, the non-dilatonic black hole has the zero horizon radius limit
where $\bar{M}=1$ (Fig.~\ref{d5-5}(a)). It is a singular solution.
There is also a lower bound on the horizon radius for the dilatonic solution.
For the minimum solution, the second derivative of the dilaton field diverges outside
the horizon.
For the charged case, the non-dilatonic solution is described by Eqs.~(\ref{ND-B})
and (\ref{ND-M}). There is the extreme solution for each charge, which has the minimum
mass. As in the RN black holes in $D=4$, the lower curve for each charge is the radius
of the inner horizon.
The graphs of the charged dilatonic solution show similar behavior of the neutral
one except for the radius of the lowest mass solution. For the minimum solution,
the second derivative of the dilaton field diverges just outside the horizon.

We can see from Fig.~\ref{d5-5}(b) that the non-dilatonic solutions has
zero temperature in the low mass limit regardless of the charge.
For the charge case, the dilatonic solution behaves like non-dilatonic one.
However, the temperature of the neutral solution raises as the
mass becomes small. Figure~\ref{d5-5}(c) shows the mass-entropy relations.

%%%%%%%%%%%%%%%%%%%%%%%%%%%%%%%%%%%%
%%%%%%%%%%%%%%%%%%%%%%%%%%%%%%%%%%%%
\section{$D=6$ Black Hole}
\label{secd=6}
%%%%%%%%%%%%%%%%%%%%%%%%%%%%%%%%%%%%
%%%%%%%%%%%%%%%%%%%%%%%%%%%%%%%%%%%%

%--figures-------------------------------------------------------------
\begin{figure}[htb]
\vspace*{-10mm}
\begin{center}
\includegraphics[width=16cm]{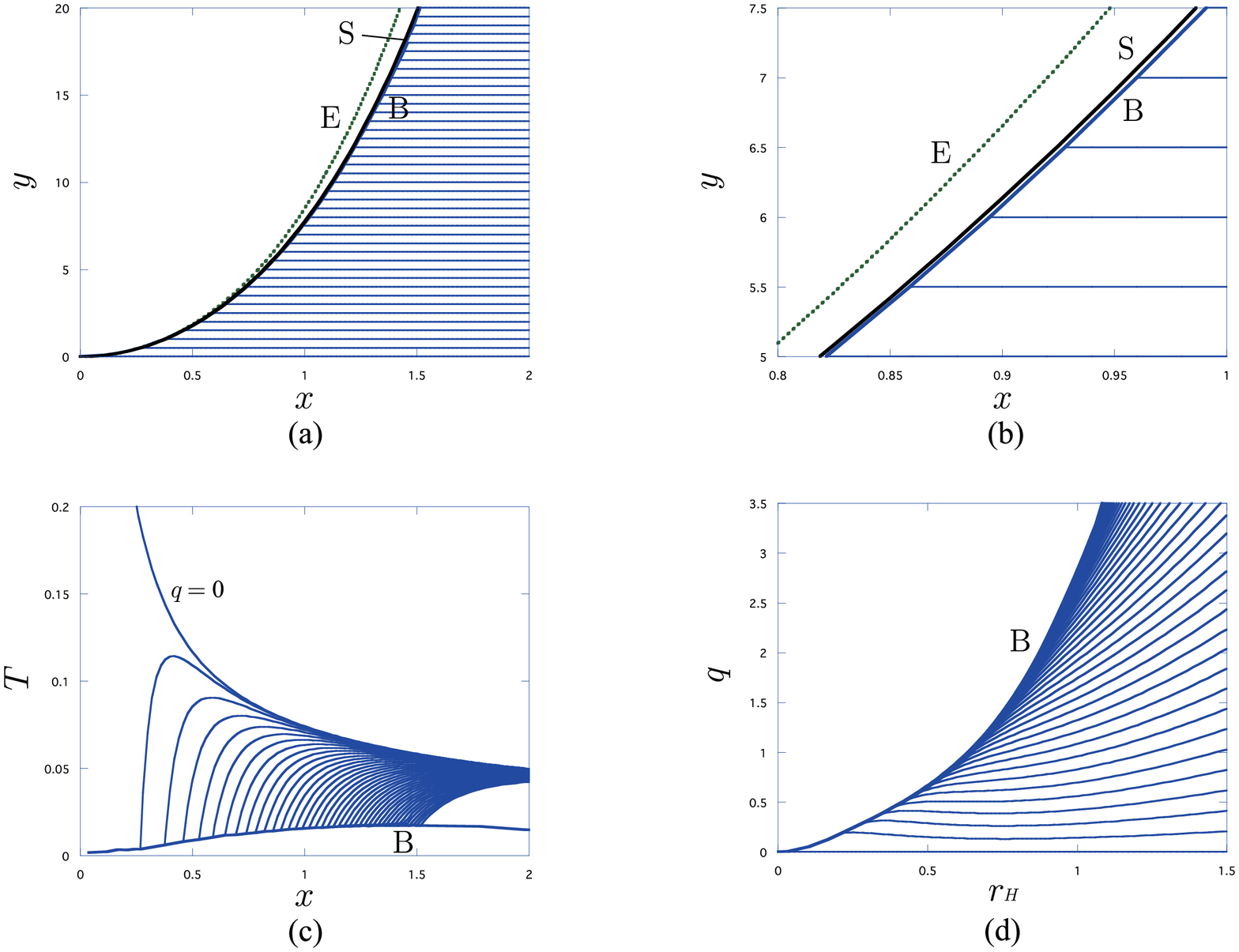}\\
\end{center}
\vspace*{-5mm}
\caption{
The parameter region where the black hole solution exists in $D=6$.
The black hole solution exists in the allowed region which is
shadowed by (blue) thin lines.
(a) The horizontal axis is $x = e^{\gamma\phi_H/2}r_H$ and the vertical
axis is $y = e^{3\gamma\phi_H/2} q$.
On the curve S the first derivative of the dilaton field
on the horizon $\phi_H'$ diverges and the solution becomes singular.
The line E represents the extreme solution.
The curve B gives
the boundary where the solutions exist. On the boundary B the second derivative of the
dilaton field diverges just outside the horizon.
(b) The magnified diagram of (a). The boundary B is separated from the
singular curve S.
(c) The allowed region in $x$ and the physical temperature $T$ of the black hole.
There are solutions above the curve B on the right side.
The solutions on the boundaries B have non-zero finite temperature.
The boundary B seems to go down to the origin.
(d) The diagram of the allowed region in terms of the physical quantity $r_H$ and $q$.
For each value of the charge $q$, there is a lower bound for the horizon radius $r_H$.
}
\label{d6-1}
\end{figure}
%--figures-------------------------------------------------------------

For $D=6$, Eq.~\p{iniphi} to determine $\phi_H'$ reduces to
\bea
&& C r_H^{10} \gamma  \Big\{4 C r_H^8 \big[4 C^2 (6 C+7) \gamma ^2+C (2 C+7)+3\big]
 -3 q^2 \big[4 C (6 C-1) \gamma ^2+2 C+1\big]\Big\} \phi_H'^2 \nn
&& ~~~-r_H \Big\{2 C \gamma ^2 \big[ 80 C^2 (C^2-C-1) r_H^{16}-12 (C-1) (4 C+1) q^2 r_H^8
 -9 q^4\big]+(2 C+1)^2 r_H^8 \big[4 C (C+3) r_H^8-3 q^2\big]\Big]\phi_H'\nn
&& ~~~-\gamma
   \Big[80 C^2 (7 C^2+2 C-2) r_H^{16}+12 (6 C^2+6 C-1) q^2 r_H^8-9 q^4\Big]=0.
\ena
The discriminant is
\bea
&& \frac{1}{4}r_H^2 \big[3 q^2-4 C (C+3) r_H^8\big]^2 \big[400 C^6 r_H^{16}+480 C^5
   r_H^{16}+24 C^4 r_H^8 (5 q^2+6 r_H^8)+8 C^3 r_H^8 (15 q^2-4 r_H^8) \nn
&&~~~~~ +C^2 (9 q^4+36 q^2 r_H^8+16 r_H^{16})+8 C r_H^8 (3
   q^2+4 r_H^8)+4 r_H^{16}\big],
\ena
which is always positive for $C>0$. Hence again there is no forbidden region from
the existence of the proper boundary condition in the $x$-$y$ diagram in Fig.~\ref{d6-1}.
There is, however,  the region where solutions cease to exist if we keep $y$ constant
and make $x$ smaller.
The curve S where $\phi'_H$ diverges and
the curve E where solution becomes extreme are expressed as
\bea
\mbox{S:~~~} y=2x^2 \sqrt{\frac{6(x^6+14x^4+108 x^2+432)}{x^4+6x^2+216}},
\label{S-d6}
\ena
\bea
\mbox{E:~~~} y=2x^2\sqrt{6(x^2+2)},
\ena
respectively. The curves S and E intersect at $(x,\;y)=(2\sqrt{5},\; 80\sqrt{33})$.

For the neutral case, there is a solution in the zero horizon limit, while there is
the lower bound for $x$ for the charged solution as shown in Fig.~\ref{d6-1}(a).
We find that the boundary is given by B where $\phi''$ diverges just outside the horizon
($r\approx r_H$). Kretschmann invariant also diverges there.
In $D=5$, the second derivative of the dilaton field diverges at $r>r_H$ on the boundary,
but there is not such behavior in $D\geq 6$.
The interval of the thin lines in the allowed region is $\Delta y =0.5$.
Figure~\ref{d6-1}(b) is the magnified diagram to distinguish B from S.
The difference between them is small.
Hence the boundary B is well approximated by Eq.~(\ref{S-d6})
for $x\lsim 2\sqrt{5}$.
{} From Fig.~\ref{d6-1}(c), the temperature does not vanish on
the boundary B but it is lower than that of closer solutions on the same curve.
On the boundary B, the temperature decreases as $y\to 0$, which corresponds to
zero charge limit, because the boundary B approaches to the extreme curve E in this limit.
However, for the neutral solution with $y=0$ exactly, the temperature
diverges in $x\to 0$ limit. This disconnected behavior is the same as that
between the Schwarzschild and the RN black hole solutions in GR.
There are solutions above the curve B on the right side
(although we do not fill in curves).
The boundary B seems to go down to the origin.
In Fig.~\ref{d6-1}(d), we display the allowed region in terms of physical quantities
$r_H$ and $q$.
The left boundary in Fig.~\ref{d6-1}(d) is B, to the left of which there is no solution.
%--figures-------------------------------------------------------------
\begin{figure}[htb]
\vspace*{-5mm}
\begin{center}
\includegraphics[width=16cm]{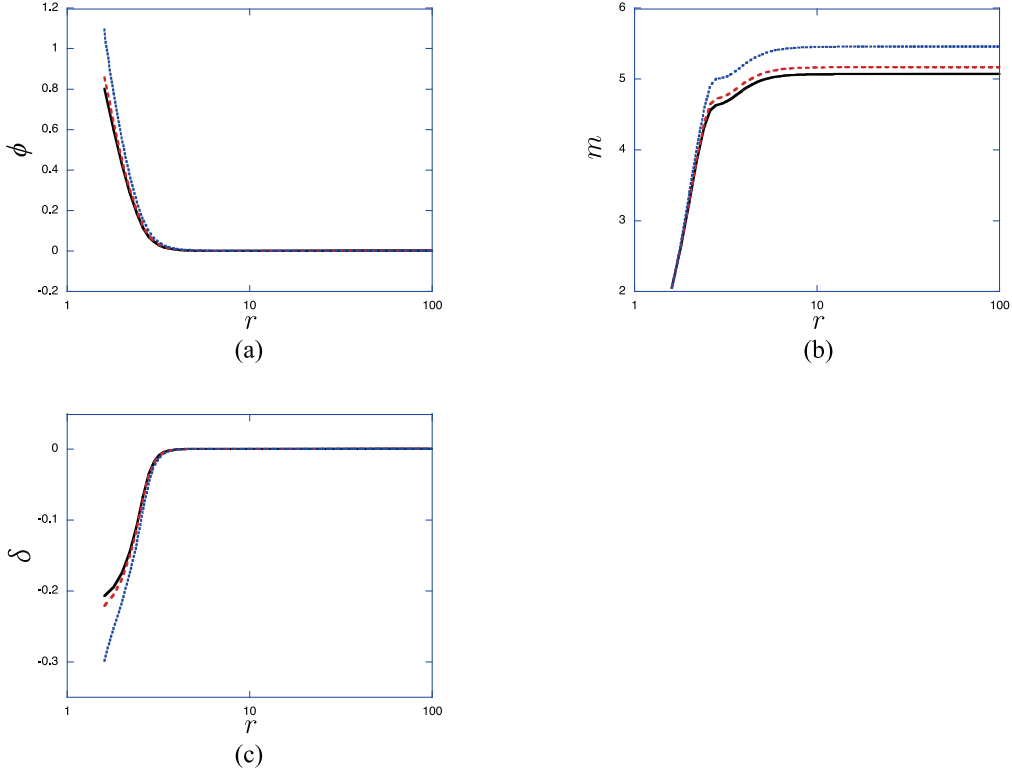}
\end{center}
\vspace*{-9mm}
\caption{Configurations of the functions; (a) the dilaton field $\phi$, (b) the mass
function $m$ and (c) the lapse function $\delta$ of the black hole solutions in $D=6$.
The horizon radii are $r_H=1.6$, and the charges are $q=0$ (solid (black) line),
 $q=5$ (dashed (red) line),  $q=10$ (dotted (blue) line).
}
\label{d6-config}
\end{figure}
%--figures-------------------------------------------------------------

Figure~\ref{d6-config} shows the configurations of the field functions $\phi$, $m$ and $\d$
for the horizon radius $r_H=1.6$ for neutral and charged cases with $q=0,5$ and 10.

We tabulate quantitative results on the physical quantities for the black hole solutions
with the charge $q=0, 5$ and 10 in Table \ref{table_d6}.
\begin{table}[h]
\begin{tabular}{@{~~~}r@{~~~}|c|@{~~~}l@{~~}|@{~~~}l@{~~}|@{~~~}l@{~~~}|@{~~~}l@{~~~}|@{~~~}l@{~~~}|@{~~~}l@{~~~}}
\hline
$q$ & ~~~$r_H$~~~ & ~~~$M$~~~ & ~~~~~~$\delta_H$ & ~~~~~~$\phi_H$ & ~~~~~$\phi_H'$ & ~~~~~~$T$
 & ~~~$S/\Sigma_1$ \\
\hline\hline
0 & 0.2 & 0.468100 & $-$0.0468327  &  ~~0.687743 & $-$0.325345  & 0.209961  &  0.170565 \\
\cline{2-8}
 & 1.7 &  5.76616 &  $-$0.205696 & ~~0.725035  &  $-$1.05676 &  0.0492691 & 14.1553 \\
\cline{2-8}
 & 2.2 & 10.3889  & $-$0.182805  & ~~0.385590  & $-$0.713747  &  0.0454031 & 29.8043  \\
\cline{2-8}
 & 4.0 & 45.6455  & $-$0.00965932  & $-$0.157190  & ~~0.00409228  & 0.0378379  &  167.850 \\
\cline{2-8}
 & 6.0 &  129.375 & ~~0.0230420 & $-$0.177248  & ~~0.0797759  & 0.0312294  & 560.017  \\
\hline
5 & 0.2 & ~~~~---  & ~~~~~~---  & ~~~~~~---  & ~~~~~~---  & ~~~~~---  & ~~~~---  \\
\cline{2-8}
 & 1.7 & 5.84751  & $-$0.215021  &  ~~0.760954 & $-$1.12794  & 0.0472207  & 13.9405  \\
\cline{2-8}
 & 2.2 & 10.4274  & $-$0.184416  & ~~0.391221  &  $-$0.723068 &   0.0450363&  29.7369 \\
\cline{2-8}
 & 4.0 & 45.6528  &  $-$0.00969279 &  $-$0.157154 &  ~~0.00400513  & 0.0378301   & 167.848  \\
\cline{2-8}
 & 6.0 & 129.377  & ~~0.0230421  & $-$0.177253  & ~~0.0797760  & 0.0312288  & 560.017  \\
\hline
10 & 0.2 & ~~~~---  & ~~~~~~---  & ~~~~~~---  & ~~~~~~---  & ~~~~~---  & ~~~~---  \\
\cline{2-8}
 & 1.7 &  6.09473 & $-$0.253558  &  ~~0.892202  & $-$1.43403  & 0.0402291  &  13.1877 \\
\cline{2-8}
 & 2.2 &  10.5425 & $-$0.189437  & ~~0.408529  &  $-$0.752338 & 0.0439171  &  29.5312 \\
\cline{2-8}
 & 4.0 &  45.6745 & $-$0.00979316  & $-$0.157046  & ~~0.00374374  &  0.0378066 &  167.842 \\
\cline{2-8}
 & 6.0 &  129.393 & ~~0.0230423  & $-$0.177268  &  ~~0.0797761 & 0.0312269  & 560.019  \\
\hline
\end{tabular}
\caption{Typical values of the physical quantities of the black hole solutions
in $D=6$.
}
\label{table_d6}
\end{table}
%--figures-------------------------------------------------------------
\begin{figure}[h]
\begin{center}
\includegraphics[width=16cm]{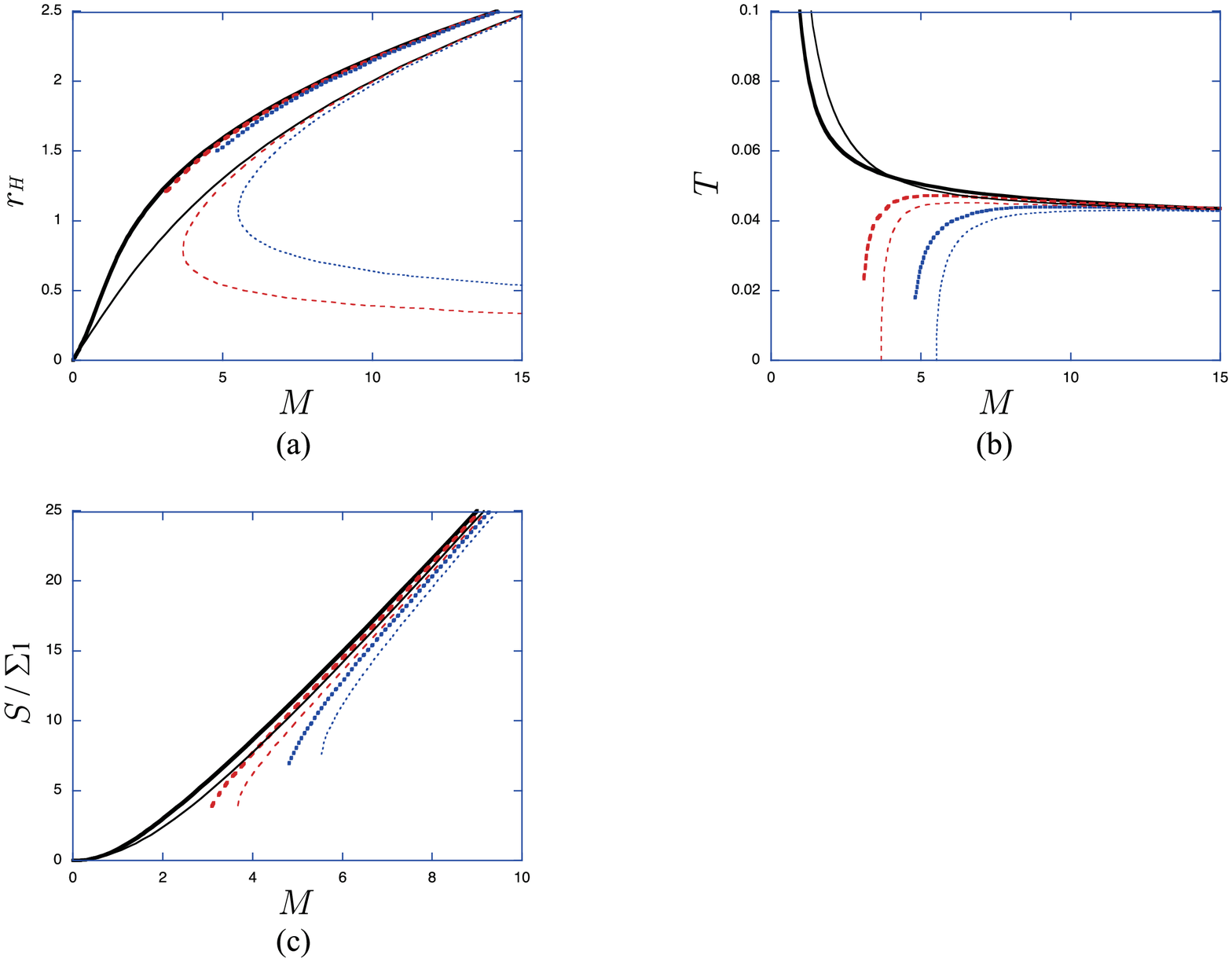}
\end{center}
\vspace*{-8mm}
\caption{The relations of physical quantities for black hole solutions in $D=6$
EGB systems.
(a) $M$-$r_H$ diagram,
(b) $M$-$T$ diagram,
(c) $M$-$S/\Sigma_1$ diagram.
Solid (black) line for $q=0$, dashed (red) line for $q=5$, and
dotted (blue) line for $q=10$.
The dilatonic solutions are given by  thick lines,
and the non-dilatonic ones by  thin lines.
}
\label{d6-3}
\end{figure}
%--figures-------------------------------------------------------------

Figure~\ref{d6-3} show the relation between physical quantities for some
fixed charges.
Figure~\ref{d6-3}(a) shows the mass and horizon radius relation.
The neutral black holes have the zero mass and zero horizon radius limit
for both the dilatonic and non-dilatonic cases. As in the lower dimensional cases
the radius of the inner horizon of the non-dilatonic solution is depicted.
In $D=6$, the horizon radius of the dilatonic solution is larger than that of the
non-dilatonic solution for the same mass, and the difference between them is remarkable.
On the other hand, the dependence of the charge in the dilatonic case is
rather mild. We can see that the $r_H$-$M$ graph of each charge traces almost the same curve.
The left end points of the curves correspond to singular solutions with the parameters on B.
Figure~\ref{d6-3}(b) shows the relation between the mass $M$ and the temperature $T$.
There appear some qualitative differences depending on whether the solution is charged or not
especially around the endpoint in the small mass limit.
Figure~\ref{d6-3}(c) shows the relation between entropy and the mass.
Compared with the $r_H$-$M$ diagram, the graphs of dilatonic solutions are
separated from each other.
Here it should be noted that the entropy is not
proportional to the horizon area in the higher curvature theory.
The expression of entropy is given by Eq.~(\ref{entropy}), and the
second term in the square bracket gives this separation.

%%%%%%%%%%%%%%%%%%%%%%%%%%%%%%%%%%%%
%%%%%%%%%%%%%%%%%%%%%%%%%%%%%%%%%%%%
\section{$D=10$ Black Hole}
\label{secd=10}
%%%%%%%%%%%%%%%%%%%%%%%%%%%%%%%%%%%%
%%%%%%%%%%%%%%%%%%%%%%%%%%%%%%%%%%%%

The qualitative properties in the $D=6$ to 10 cases are almost the same. However,
since $D=10$ is the critical dimension in superstring theories and important
for applications, here we show the summary and some diagrams
of the black hole solutions in $D=10$.

For $D=10$,  Eq.~\p{iniphi} to determine $\phi_H'$ reduces to
\bea
&& C r_H^{18} \c \Big\{8 C r_H^{16} \bigl[24 C^2 (70 C+19) \c^2+90 C^2+57C+7\bigr]
-7 q^2 \big[8 C (14 C-1) \c^2+6 C+1\big]\Big\} \phi_H'^2 \nn
&& ~~~~~
-r_H \Big\{8 C r_H^{32} \bigl[432C^2 (15 C^2-C-1) \c^2+540 C^3+108C^2+99C+7\bigr]
\nn
&&~~~~~~~~~~
- 7 q^2 r_H^{16} \bigl[16C(C-1)(12 C+1) \c^2 +(6 C+1)^2\bigr]-98 C q^4 \gamma ^2\Big\}\phi_H'
\nn
&& ~~~~~
-\c \big[576 C^2 (99 C^2+6 C-4) r_H^{32}+56 (18 C^2+6 C-1) q^2 r_H^{16}-49 q^4\big]
=0.
\ena
The discriminant is
\bea
&& \frac{1}{4} r_H^2 \big[7 q^2-8 C (15 C+7) r_H^{16}\big]^2
\big[46656 C^6 r_H^{32}+34560 C^5 r_H^{32}+432 C^4 (7 q^2 r_H^{16}+20 r_H^{32}) \nn
&& ~~~~~
+192 C^3 (7 q^2 r_H^{16}+9 r_H^{32})+C^2 (7 q^2+24 r_H^{16})^2+8 C (7 q^2
   r_H^{16}+12 r_H^{32})+4 r_H^{32}\big],
\ena
which is always positive. Hence again there is no forbidden region in the parameter
space from this restriction.

\begin{figure}
\vspace*{-10mm}
\begin{center}
\includegraphics[width=16cm]{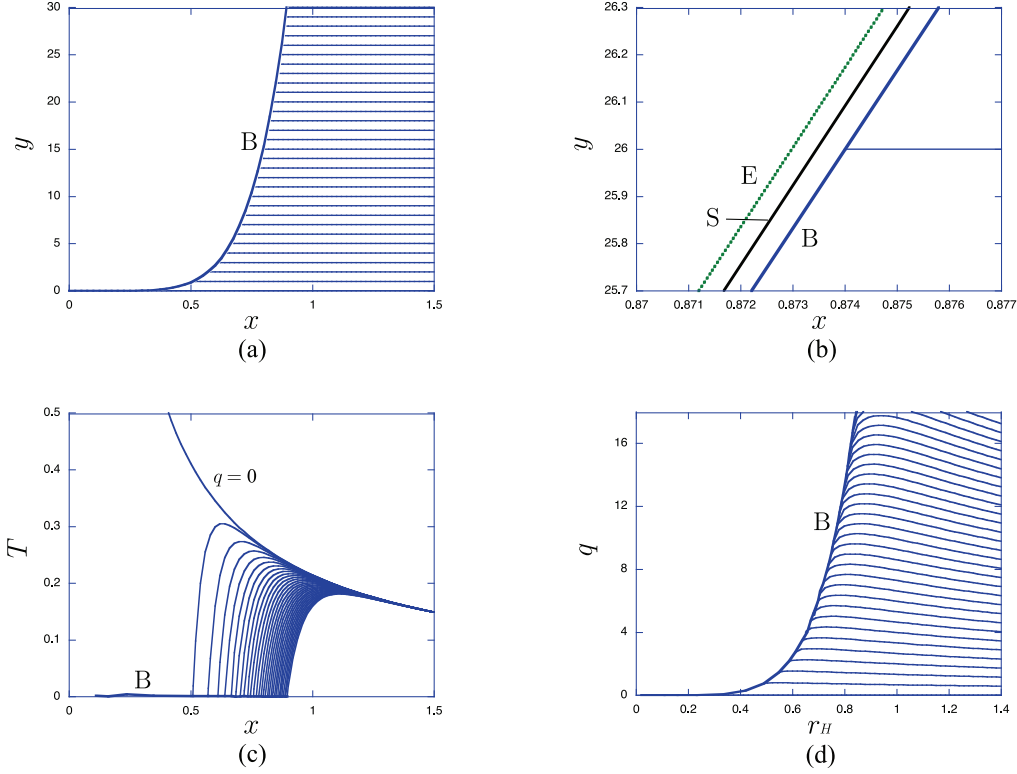}\\
\end{center}
\caption{The parameter region where the black hole solution exists in $D=10$.
The black hole solution exists in the allowed region which is
shadowed by (blue) thin lines.
(a) The horizontal axis is $x = e^{\gamma\phi_H/2}r_H$ and the vertical
axis is $y = e^{4\gamma\phi_H} q$.
The curve B gives the boundary where the solutions exist.
On the boundary B the second derivative of the dilaton field diverges just outside the horizon.
(b) The magnified diagram of (a).
There are small differences between the boundary B,
the singular curve S, and the extreme curve E.
(c) Allowed region in $x$ and the physical temperature $T$ of the black hole.
There are solutions also in the lower-right region.
The solutions on the boundaries B have non-zero finite temperature.
The boundary B seems to go down to the origin.
(d) The diagram of the allowed region in terms of the physical quantity $r_H$ and $q$.
For each value of the charge $q$, there is a lower bound for the horizon radius $r_H$.
}
\label{d10-1}
\end{figure}

In Fig.~\ref{d10-1}, we give the allowed region for the dilatonic black holes.
The interval of the thin lines in the allowed region is $\Delta y =1$.
In Fig.~\ref{d10-1}(a), the extremal curve E and the singular curve S where $\phi'_H$
diverges overlap with the boundary B, which are expressed as
\bea
\mbox{S:~~~} y=4x^6 \sqrt{\frac{7(x^6+114x^4+5712 x^2+164640)}{x^4+56x^2+5488}},
\ena
\bea
\mbox{E:~~~} y=4x^6\sqrt{7(x^2+30)},
\ena
respectively.
The curves S and E intersect at $(x,\;y)= (2\sqrt{13},\; 162432\sqrt{574})$.
Figure~\ref{d10-1}(b) is the magnified diagram of (a). We can see that there are actually
small differences between these curves.
On the boundary B, the second derivative of  the dilaton field diverges just outside the horizon.
Figure~\ref{d10-1}(c) shows the $T$-$x$ diagram.
There are solutions between the $q=0$ curve and the boundary B.
On the boundary B the temperature is not zero.
In Fig.~\ref{d10-1}(d), we display the allowed region in terms of $x$ and $y$.
Because the value of $\phi_H (<0.3)$ is small for all ranges, there is not much difference from
Fig.~\ref{d10-1}(a).

%--figures-------------------------------------------------------------
%--figures-------------------------------------------------------------

The configurations of the field functions $\phi$, $m$ and $\d$ are qualitatively
similar to $D=6$ case, and we omit the figures here.
Instead, we give quantitative results on the typical physical quantities for
the black hole solutions with the charge $q=0, \: 5$ and 10 in Table \ref{table_d10}.
\begin{table}[h]
\begin{tabular}{@{~~~}r@{~~~}|c|@{~~~}l@{~~}|@{~~~}l@{~~}|@{~~~}l@{~~~}|@{~~~}l@{~~~}|@{~~~}l@{~~~}|@{~~~}l@{~~~}}
\hline
$q$ & ~~~$r_H$~~~ & ~~~~$M$~~~ & ~~~~~~$\delta_H$ & ~~~~~~$\phi_H$ & ~~~~~$\phi_H'$ & ~~~~~~$T$
 & ~~~$S/\Sigma_1$ \\
\hline\hline
0 & 0.2 & 0.0066645 & $-$0.00168778  &  0.0355222 & $-$0.0319461  & 0.996737  &  0.00176109 \\
\cline{2-8}
 & 1.7 &  6.48277 &  $-$0.0125900 & 0.163543  &  $-$0.133113 &  0.253318 & 6.80565 \\
\cline{2-8}
 & 2.2 & 276.002  & $-$0.0368148  & 0.318372  & $-$0.276744  &  0.125172 & 593.831  \\
\cline{2-8}
 & 4.0 & 1027.74  & $-$0.0497577  & 0.361146  & $-$0.333154  & 0.100354  &  2787.35 \\
\cline{2-8}
 & 6.0 &  26445.5 & $-$0.0693803  & 0.285279  & $-$0.335072  & 0.0646700  & 115826  \\
\hline
5 & 0.2 & ~~~~---  & ~~~~~~---  & ~~~~~---  & ~~~~~---  & ~~~~~---  & ~~~~~---  \\
\cline{2-8}
 & 1.7 & 7.06843  & $-$0.0138779  &  0.170767 & $-$0.147951  & 0.221395  & 6.80611  \\
\cline{2-8}
 & 2.2 & 276.008  & $-$0.0368154  & 0.318375  &  $-$0.276748 &   0.125170 &  593.831 \\
\cline{2-8}
 & 4.0 & 1027.74  &  $-$0.0497577 &  0.361147 &  $-$0.333155  & 0.100354   & 2787.35  \\
\cline{2-8}
 & 6.0 & 26445.5  & $-$0.0693803  & 0.285279  & $-$0.335072  & 0.0646700  & 115826  \\
\hline
10 & 0.2 & ~~~~---  & ~~~~~~---  & ~~~~~---  & ~~~~~---  & ~~~~~---  & ~~~~~---  \\
\cline{2-8}
 & 1.7 &  8.76069 & $-$0.0207554  &  0.193860  & $-$0.236732  & 0.123396  &  6.70390 \\
\cline{2-8}
 & 2.2 &  276.018 & $-$0.0368170  & 0.318383  &  $-$0.276759 & 0.125164  &  593.828 \\
\cline{2-8}
 & 4.0 &  1027.74 & $-$0.0497578  & 0.361147  & $-$0.333155  &  0.100354 &  2787.35 \\
\cline{2-8}
 & 6.0 &  26445.5 & $-$0.0693803  & 0.285279  &  $-$0.335072 & 0.0646700  & 115826  \\
\hline
\end{tabular}
\caption{Typical values of the physical quantities of the black hole solutions
in $D=10$.
}
\label{table_d10}
\end{table}

\begin{figure}[h]
\vspace*{-5mm}
\begin{center}
\includegraphics[width=16cm]{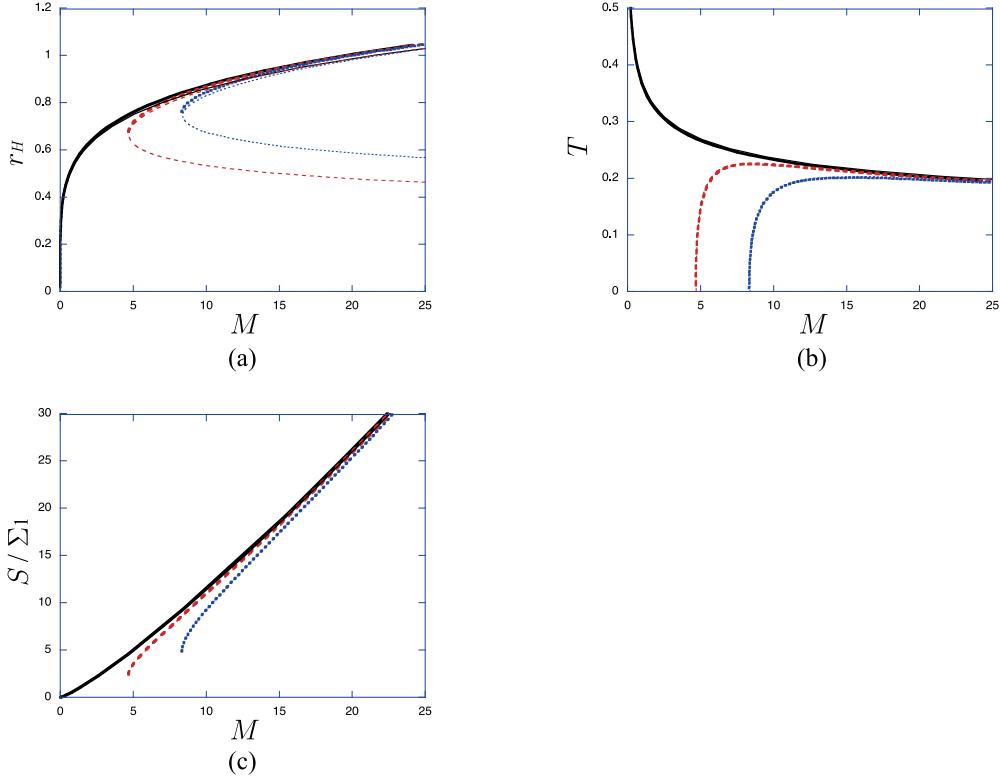}
\end{center}
\vspace*{-3mm}
\caption{The relations of various physical quantities for black hole solutions in
$D=10$ dilatonic EGB systems.
(a) $M$-$r_H$ diagram,
%(b) $M$-$r_{s}$ diagram,
(b) $M$-$T$ diagram,
(c) $M$-$S/\Sigma_1$ diagram.
Solid (black) line for $q=0$, dashed (red) line for $q=5$, and dotted (blue)
line for $q=10$.
The dilatonic solutions are given by  thick lines,
and the non-dilatonic ones by thin lines.
}
\label{d10-3}
\end{figure}

In Fig.~\ref{d10-3}, we give the relations of physical quantities.
There is not much difference between the dilatonic and the non-dilatonic solutions
for these values of charge.
This seems to be the tendency for higher-dimensional solutions.
%--figures-------------------------------------------------------------
%--figures-------------------------------------------------------------

%--figures-------------------------------------------------------------
%\begin{figure}
%\vspace*{-5mm}
%\begin{center}
%\includegraphics[width=16cm]{figures/fig_config_d10.eps}
%\end{center}
%\vspace*{-9mm}
%\caption{Configurations of the functions (a) $\phi$,  (b) $m$ and  (c) $\delta$ of
%the black hole solutions in $D=10$.
%The horizon radii are $r_H=0.8$, and the charges are $q=0$ (black solid line),
% $q=5$ (red dashed line),  $q=10$ (blue dotted line).
%}
%\label{d10-config}
%\end{figure}
%--figures-------------------------------------------------------------

%%%%%%%%%%%%%%%%%%%%%%%%%%%%%%%%%%%%
%%%%%%%%%%%%%%%%%%%%%%%%%%%%%%%%%%%%
\section{Conclusions %and Discussions
}
\label{CD}
%%%%%%%%%%%%%%%%%%%%%%%%%%%%%%%%%%%%
%%%%%%%%%%%%%%%%%%%%%%%%%%%%%%%%%%%%

In this paper we have studied asymptotically flat charged black hole solutions
in the dilatonic EGB theory in various dimensions. The theory is the low-energy
effective theory of the heterotic string. %, and hence a important system.
The spacetime is assumed to be static and spherically symmetric.
The system of the field equations is so complex that it is difficult to
obtain an analytical solution. Hence we investigate the numerical solutions.
The system of the field equations has some symmetries which are helpful in the analysis.
The results are given for $D=4, 5,6$ and 10, and we did not discuss other dimensions
7 to 9 simply because we expect that behaviors are qualitatively the same as
those presented here from our earlier study of the system.

We have found that there is the forbidden region on the parameter plane spanned by $x$ and
$y$ which are the ``scaled horizon radius" and  the ``scaled charge", respectively, in $D=4$.
Besides it, there are some boundaries (and important curves) of the allowed region such as the
singular curve S where $\phi'_H$ diverges, the extreme curve E where the solution becomes
extreme, and B$_i$ ($i=1,~2,~3$) where the field functions and/or its derivatives diverge.
The forms of the allowed parameter regions are
different depending on the dimension, 4, 5, and 6 -- 10.

We have also studied the thermodynamical quantities. There is no extreme black hole solution
with $T=0$ that can be obtained by taking the limit of the non-extreme solutions
within the parameter range we chose. In the higher curvature theory,
physical entropy is defined by Iyer and Wald~\cite{Wald}. Although there is a parameter
region where the radius (or the area) of the black hole horizon in the dilatonic theory
is smaller than that in the non-dilatonic theory, entropy in the  dilatonic theory
is always larger than that in the non-dilatonic theory.
Since the dilatonic and non-dilatonic solutions are the solutions in different
theories, it is difficult to compare their thermodynamical
stability and quantum transition between them in the thermodynamical
sense as long as $\gamma\ne 0$. There would be, however, some physical meaning
in this magnitude relationship of entropy.

It is noted again that our analysis includes the higher order term of the dilaton field
which is not in our previous works \cite{GOT1,GOT2,OT3,OT4,OT5}.
To make its effects clear, we have studied the solutions for both cases $a=0$ and $a=1$
in Eq.~\p{action1} in the Appendix by focusing on the neutral solutions.
The qualitative properties and the relations such as $M$-$r_H$, $M$-$T$ and $M$-$S/\Sigma_1$
are quite similar in each case except for $D=5$.
In the $D=5$ theory without the higher order term,
the black hole solution with infinitesimal size exists towards $r_H=0$
while there is a lower bound on the horizon radius $r_H$
in the theory with the higher order term.
For the lower bound solution, the second derivative of the dilaton field
diverges at some radius outside the horizon, and the spacetime becomes singular.
The existence of the lower bound of this type is confirmed also in the charged
solution with $y\lesssim 12$ ($q\lesssim 0.1$).

By our analysis it is found that
the properties of the black hole solutions strongly depend on the dimension,
charge, existence of the dilaton field. Hence both the detailed analyses of the
individual systems and the investigations from the systematic point of view
are important.

There still remain some questions in our work.
The first one is to determine precisely where the real boundary of the allowed region
B (or B$_2$) is. It is difficult to determine it due to the fine structure of the system near
the boundary and the numerical accuracy.
The second is concerned with the fact that the temperature of the black hole solutions
on some boundary remains non-zero and finite. This means that the evaporation does not
stop there and the black hole still evolves to a naked singularity or something unknown.
This gives a very interesting puzzle that the singularity may be really formed after
evaporation process and deserves further study.
%We should include the higher order terms of the effective theory there.
%However, it is difficult to give definite answer to it in the present analysis.

Another interesting future work is to study charged AdS black holes for application
to AdS/CFT correspondence. We expect that this class of solutions exists for $k=0$~\cite{GOT2}.
We leave these problems for future works.

\section*{Acknowledgements}
This work was supported in part by the Grant-in-Aid for
Scientific Research Fund of the JSPS (C) Grant No. 24540290, (C) Grant No. 22540293and,
and (A) Grant No. 22244030.

%%%%%%%%%%%%%%%%%%%%%%%%%%%%%%%%%%%%%%%%%%%%%%%%%%%%%%%%%%%
\appendix

%%%%%%%%%%%%%%%%%%%%%%%%%%%%%%%%%%%%%%%
%%%%%%%%%%%%%%%%%%%%%%%%%%%%%%%%%%%%%%%
\section{The effects of the higher order term of the dilaton field for the neutral solutions}
%%%%%%%%%%%%%%%%%%%%%%%%%%%%%%%%%%%%%%%%

In our previous work on the black holes in the dilatonic EGB theory
\cite{GOT1,GOT2,OT3,OT4,OT5}, we examined the neutral solutions without the higher order
derivative term of the dilaton field ($a=0$ in Eq.~\p{action1}).
There is, however,  such a term in general.
Hence it is significant to investigate if there are any differences between the cases of
$a=0$ and $a=1$.
The neutral solutions for $a=1$ can be obtained by putting $q=0$ in our model.
One can find the result by looking at the curves with $q=0$ in Figs.~\ref{d4-1}-\ref{d4-3}
for $D=4$.
Comparing them with the  $a=0$ case in our previous papers, we find that the qualitative
properties and the relations such as $M$-$r_H$, $M$-$T$ and $M$-$S/\Sigma_1$ are quite similar.
This is true also in other dimensions except for five.
Hence let us discuss $D=5$ case in more detail here.

%--figures-------------------------------------------------------------
\begin{figure}[h]
\vspace*{-10mm}
\begin{center}
\includegraphics[width=16cm]{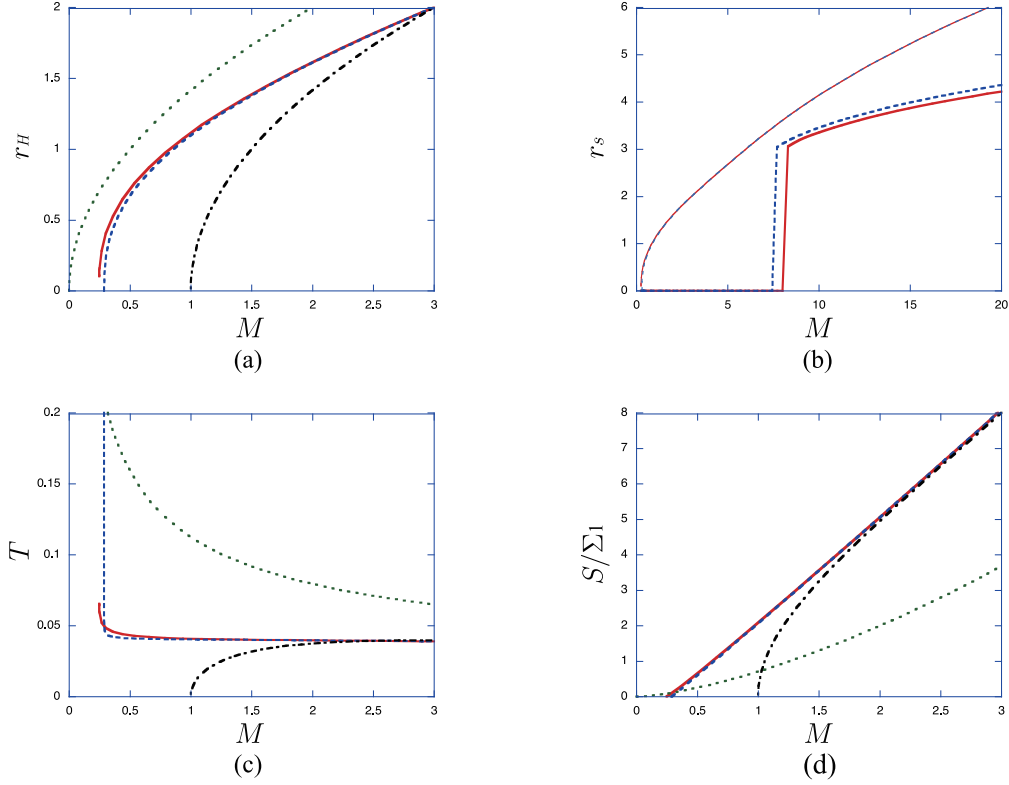}
\end{center}
\vspace*{-5mm}
\caption{Difference between $a=1$ and $a=0$ for neutral black hole solutions
in the $D=5$ dilatonic EGB system with $\c=1/2$.
(a) $M$-$r_H$ diagram,
(b) $M$-$r_s$ diagram (The thin lines show the horizon radius),
(c) $M$-$T$ diagram,
(d) $M$-$S/\Sigma_1$ diagram.
The solid (red) line is for $a=1$, the dashed (blue) line for $a=0$,
the dotted (green) line for Tangherini solution (higher-dimensional generalization
of the Schwarzschild solution without the GB term), and
the dotted-dashed (black) line for the BD solution (in non-dilatonic EGB theory).
}
\label{d5-1}
\end{figure}
%--figures-------------------------------------------------------------

\begin{figure}[htb]
\vspace*{-5mm}
\begin{center}
\includegraphics[width=7cm]{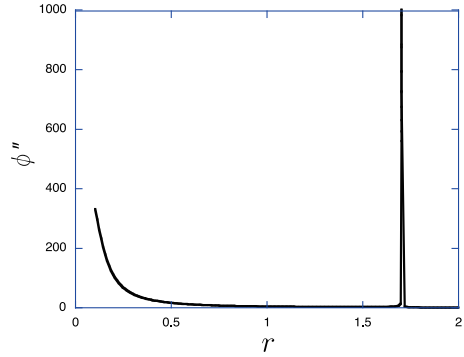}
\end{center}
\vspace*{-10mm}
\caption{
Configuration of $\phi''$ for parameters $q=0, ~r_H=0.103555$ just before touching
the lower bound in $D=5$.
The peak is at $r=1.702$. This behavior is basically the same for the charged solutions
with $y<12$ ($q <0.1$).
}
\label{d5-2}
\end{figure}

Figure~\ref{d5-1} displays the relations between physical quantities for neutral solutions.
The black hole solutions in the theory with the GB term  have the non-zero lower bound on
their mass.
Compared with the non-dilatonic BD solution,
the mass of the lower bound in the dilatonic solutions is smaller,
and the one for $a=1$ is slightly smaller than that for $a=0$ (Fig.~\ref{d5-1}(a)).
We also find that for the $a=0$ case
the black hole solution with infinitesimal size exists towards $r_H=0$
while there is a lower bound on the horizon radius $r_H$ for $a=1$.
This bound corresponds to the boundary B in Figs.~\ref{d5-3}(a) and \ref{d5-4}(a).
With the parameters around the lower bound B
the second derivative of the dilaton field grows suddenly at some radius $r>r_H$.
This behavior is shown in Fig.~\ref{d5-2} for the parameters
just before touching the boundary.
It is expected that $\phi''$ should diverge on the boundary.
Since the  Kretschmann invariant also diverges at that radius,
the solution of the lower bound is singular.
If we compare black holes with the same mass, the horizon radius for $a=1$ is
a little larger than that for $a=0$.
As a general tendency, values of the dilaton field $\phi$ itself and its variation $\phi'$
are larger in magnitude for $a=1$ than those for $a=0$.
For the large black holes ($r_H \gtrsim 1$), the differences are indistinguishable.
%--figures-------------------------------------------------------------
%--figures-------------------------------------------------------------

Figure~\ref{d5-1}(b) shows the radius of the curvature singularity (thick curves) inside
the black hole horizon (of which radius is depicted by thin curves).
For $M\lesssim 7.5$, the curvature singularity is located at the center of the
black hole. For the large mass $M\gtrsim 7.5$, however, the singularity is expanded to
a finite radius $r_s>0$. We called it a fat singularity~\cite{OT5}.
Even if the higher order term of the dilaton field is included, this tendency is unchanged.

The temperature $T$ is slightly higher in $a=1$ than in $a=0$
case for $M\gtrsim 0.6$,
while the order changes near the minimum mass (Fig.~\ref{d5-1}(c)).
Although it might appear that the difference is very small in $D=5$, the difference is
further smaller in other dimensions.
In addition, the temperature of the minimum size solution is finite,
hence a small black hole for $a=1$ may evolve to this solution through the evaporating
process, and the naked singularity might appear at non-zero radius.
Such behavior cannot be seen for higher dimensions even if the higher order
term is included.
Figure~\ref{d5-1}(d) is the entropy vs. mass diagram.

%%%%%%%%%%%%%%%%%%%%%%%%%%%%%%%%%%%%%%%%%

\end{document}